\documentclass[11pt]{article}
\usepackage[final]{acl}

\usepackage{times}
\usepackage{latexsym}
\usepackage[T1]{fontenc}
\usepackage[utf8]{inputenc}
\usepackage{microtype}
\usepackage{inconsolata}
\usepackage{graphicx}
\usepackage[table]{xcolor}
\usepackage{array}
\usepackage{pifont}

\usepackage{enumitem}
\setlist[itemize]{noitemsep, nosep}
\usepackage{booktabs}
\usepackage{amssymb}
\usepackage{amsthm}
\usepackage{amsmath}
\usepackage{acro}
\usepackage{multirow}
\usepackage{siunitx}
\usepackage{tabularx}
\usepackage{ragged2e}
\usepackage{soul}
\usepackage{orcidlink}
\usepackage{censor}
\usepackage[role = rolecolor, grid = gridcolor]{credits}

\newcommand{\explink}{\url{https://github.com/dimits-ts/interventions}}
\newcommand{\datasetlink}{\url{https://github.com/dimits-ts/facilitation-dataset}}
\newcommand{\surveydatalink}{\url{https://github.com/dimits-ts/annotation}}

\acsetup{make-links}
\DeclareAcronym{nlp}{short = NLP, long = Natural Language Processing}
\DeclareAcronym{nlu}{short = NLU, long = Natural Language Understanding}
\DeclareAcronym{ml}{short = ML, long = Machine Learning}
\DeclareAcronym{rq}{short = RQ, long = Research Question}
\DeclareAcronym{pefk}{short = PEFKI, long = Prosocial and Effective Facilitation in Konversational Interactions}
\DeclareAcronym{eo}{short = EncOn, long = Encoder-Only}
\DeclareAcronym{iaa}{short = IAA, long = Inter-Annotator Agreement}
\DeclareAcronym{umod}{short = UMOD, long = User Moderation}
\DeclareAcronym{iq2}{short = IQ2, long = Intelligence Squared}
\DeclareAcronym{ceri}{short = CeRI, long = Cornell eRulemaking Initiative}
\DeclareAcronym{whow}{short = WHoW, long = Why-How-Who}
\DeclareAcronym{awry}{short = AWRY, long = Convs. Gone Awry}
\DeclareAcronym{cmv}{short = CMV II, long = Change My View}
\DeclareAcronym{wt}{short = WT, long = WikiTactics}
\DeclareAcronym{cot}{short = CoT, long = Chain of Thought}
\DeclareAcronym{gpl}{short=GPLv3, long=GNU Public License version 3-or-later}

\newcommand{\cmark}{\ding{51}}
\newcommand{\xmark}{\ding{55}}

\definecolor{rolecolor}{RGB}{212,175,55}
\definecolor{gridcolor}{RGB}{255, 255, 255}

\newcommand{\KKorcid}{\orcidlink{0000-0002-9349-9554}}
\newcommand{\DTorcid}{\orcidlink{0000-0002-5675-3939}}
\newcommand{\JPorcid}{\orcidlink{0000-0001-9188-7425}}

\theoremstyle{definition}
\newtheorem{definition}{Definition}

\theoremstyle{plain}
\newtheorem{finding}{Finding}
\newtheorem*{researchquestion}{RQ}

\author{Dimitris Tsirmpas\DTorcid \and Katerina Korre\KKorcid \and John Pavlopoulos\JPorcid\\
	Athens University of Economics and Business and \\
	Archimedes, Athena Research Center, Greece \\
	\texttt{dim.tsirmpas@aueb.gr}, \texttt{k.korre@athenarc.gr}, \texttt{ipavlopoulos@aueb.gr} \\
}

\title{To Facilitate or Not to Facilitate: Human and LLM Facilitator Tendencies in Online Discussions}

\graphicspath{{../graphs} {graphs}}

\begin{document}
	\maketitle
	\begin{abstract}
		Automating facilitation in online discussions is a long-standing social concern given the increasing time we spend on online spaces and the failure of content moderation approaches. While studies have been conducted on how to facilitate, none have answered the essential question of when to do so. A potential answer is using LLMs, which ostensibly make automated, large-scale intervention increasingly feasible. In this study, we examine when LLMs decide to facilitate by defining what facilitation is, observing when humans decide to facilitate, and comparing their decisions with those made by LLMs. To this end, we create \ac{pefk}, a corpus standardizing and aggregating all relevant facilitation datasets. We are the first to run a survey on facilitation timing, which we execute using expert facilitative participants and LLM-as-a-judge models. We discover that while humans are more cautious, LLMs are excessively eager to facilitate, although both are more certain when judging that facilitation is not needed.	We then investigate whether this behavior can be corrected using alternative setups for LLMs and training ModernBERT classifiers on established datasets, finding that the latter perform more reliably than the former, although current datasets result in relatively low performance.
	\end{abstract}
	
	\acresetall
	
	\section{Introduction}
	\label{sec:intro}
	
	\begin{figure}[!t]
		\includegraphics[width=0.9\columnwidth]{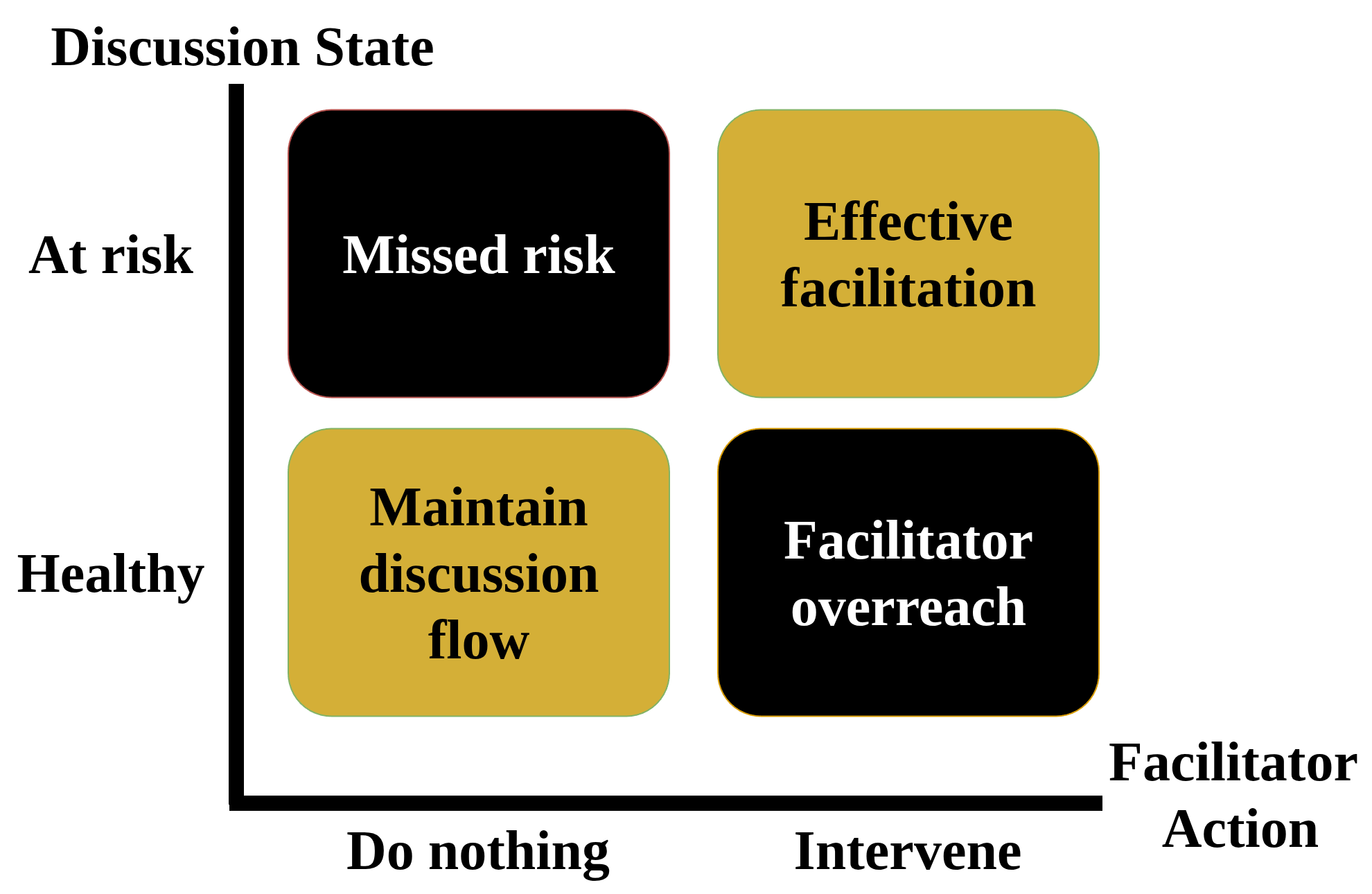}
		\caption{At any point in a discussion, the facilitator must decide whether to intervene or not. Not intervening when the situation calls for it may lead to the discussion worsening (e.g., topic derailment, escalation, missed participation). On the other hand, intervening when the discussion is already going well may lead to irritation among participants.}
		\label{fig:intro}
	\end{figure}
	
	Given the scale of social media networks and online spaces, facilitation needs to be automated to keep up with the amount of posted content~\citep{small-polis-llm, Burton2024}. So far, platforms have largely ignored the issue, preferring automated flag-and-remove mechanisms (``content moderation'')~\citep{korre-etal-2025-evaluation}, which are not sufficient~\citep{cresci_2022,trujillo_2022_make_reddit}. While recent work has undertaken the task of choosing \textit{how} to facilitate~\citep{chen-etal-2025-whow, schroeder-etal-2024-fora, ceri}, there has been little interest in \textit{when} to do so \citep{falk-etal-2021-predicting}.
	
	Deciding when to intervene in a discussion is a matter of great importance. Intervening too early or too frequently may lead to users expressing frustration at being constantly interrupted, while intervening too late may lead to a significant escalation and derailment of the discussion (Fig.~\ref{fig:intro}, \citet{korre-etal-2025-evaluation, gao-etal-2025-moderation}). Due to the sheer scale of modern online spaces automation is needed, which the field of \ac{nlp} has attempted to implement using models that determine when an intervention is necessary. \ac{eo} classifiers were used to this end~\citep{falk-etal-2021-predicting}, although recently LLMs have been championed as the eventual solution to online facilitation~\citep{korre-etal-2025-evaluation, small-polis-llm}. In this study, we focus on the following \ac{rq}:
	
	\begin{researchquestion}
		\label{rq:aberrant-llms}
		How and why do LLMs differ with humans when deciding when to facilitate?
	\end{researchquestion}
	
	Since humans are generally competent at facilitation \citep{umod} especially when trained \citep{schroeder-etal-2024-fora, park2016supporting}, we would expect that automated facilitation models, if competent, would roughly replicate human tendencies. This is the main assumption made in this study, as there currently is not a reliable and standardized methodology for evaluating discussions through metrics and annotation \citep{korre-etal-2025-evaluation}. The main research question can be answered through the following sub-questions:
	
	\paragraph{What Constitutes a Facilitative Intervention?} There is currently no consensus on how to define a facilitative intervention. By examining relevant literature, we find three common definitions implicitly used in prior work; \emph{escalation, professional facilitation}, and \emph{facilitative comments} (\S\ref{sec:related:interventions}). 
	
	\paragraph{Where Do Humans Facilitate?} While the majority of content where facilitation is needed is posted on online, written discussions, recent work predominantly focuses on \emph{domain-specific oral discussion} scenarios (e.g., radio debates).	Since datasets for discussion facilitation are generally few, small, and straddle radically different domains (e.g., debates, forums, deliberations), we gather all datasets relating to any of the three possible definitions of the term and compile the ``\ac{pefk}'', a standardized dataset for discussion facilitation with a unified data and label schema (\S\ref{sec:dataset}).
	
	\paragraph{How Do Humans Facilitate?} We create and execute a survey task designed to gauge human facilitative patterns by sampling 1224 discussion chunks from our dataset, and asking 10 expert facilitative participants whether they would intervene as facilitators. We find that human participants tend to take a cautious approach to facilitation, generally preferring not to intervene and expressing greater confidence in not doing so (Finding~\ref{finding:human-facilitate}). Additionally, human participants may pay attention to different spans of a discussion, which can lead them to choose different courses of action regarding facilitation.
	
	\paragraph{How Does LLM facilitation differ?} We repeat the survey with six open-source LLMs. Unlike humans, LLMs tend to over-intervene and rely more heavily on positive reinforcement. However, compared to humans, they may be more consistent between themselves, hinting that they are likely focusing on similar spans of the discussion. We also find that, much like humans, determining when not to facilitate is easy--the opposite is not (Findings~\ref{finding:llm-facilitate-pefk},\ref{finding:no-fac-easy}).
	
	\paragraph{How Can We Fix LLM Facilitation?}
	We investigate whether aggregating all currently available datasets can support either the training of classifiers or the use of off-the-shelf LLMs for a simpler and more commonly studied task in the literature: predicting whether a facilitator would intervene, rather than whether a facilitative comment is needed (\S\ref{sec:facilitator}). Our results show that LLMs are less reliable than ModernBERT~\citep{warner-etal-2025-smarter} classifiers, although their performance depends heavily on the model used (Finding~\ref{finding:llms-may-facilitate}).
	
	Overall, our study offers the following contributions: We bridge differences between prior work on (i) what a facilitation intervention is, (ii) oral and written discussions (1), by compiling the largest standardized facilitation dataset so far (2). Through the execution of an extensive survey performed by experts on facilitation (3), we show that the key difference between humans and LLMs rests in how frequently they deem facilitation to be necessary (4). We are the first to evaluate LLMs for facilitation timing prediction, finding that they are highly inconsistent while \ac{eo} models offer a more reliable alternative, though both approaches remain constrained by the limited current understanding of facilitation within \ac{nlp} (5). 
	
	Ultimately, achieving reliable automated facilitation requires moving beyond surface signals and using comprehensive, human-informed data that captures the true complexity of human facilitation. We release the code for the creation of our dataset and all our experiments under a \ac{gpl} license.\footnote{Dataset: \datasetlink, Experiments: \explink, Survey Data: \surveydatalink}

	\section{Related Work and Background}
	\label{sec:related}
	
	\subsection{What Are Facilitative Interventions?}
	\label{sec:related:interventions}
	
	Facilitation (or conversational moderation) refers to the active participation of people in order to improve discussions~\citep{korre-etal-2025-evaluation}. Relevant work assumes that facilitation is conducted by specially trained staff~\citep{schroeder-etal-2024-fora, ceri, chen-etal-2025-whow}, or users appointed by the community ~\cite{schaffner_community_guidelines}. Studying relevant literature on facilitation yields multiple definitions for what a facilitative intervention is, which can be categorized to three main definitions:
	
	\begin{definition}[Escalation]
		\label{def:escalation}
		A facilitator needs to intervene at the precise point at which, without facilitation, the discussion would derail or escalate.
	\end{definition}
	
	Escalation is usually measured as the point where the participants request arbitration (e.g., escalate the matter to a site moderator in Wikipedia~\cite{wikidisputes, cmv_awry2}). This definition is quite appealing for \ac{nlp} researchers as the labels are easy to acquire since a ground truth exists, and the existence of a single, largely unambiguous ground-truth is ideal for the training and evaluation of any \ac{ml}-based classifier. However, a facilitator has multiple duties that do not directly relate to preventing escalation such as summarizing discussions~\cite{small-polis-llm}, managing participation~\cite{chen-etal-2025-whow, schroeder-etal-2024-fora}, and providing important information for new members~\cite{ceri}. Furthermore, an unambiguous signal that a discussion has escalated is not often provided by platforms (Table~\ref{tab:datasets}). Thus, this definition is not considered in this study, but is included in our dataset (\S\ref{sec:dataset}).
	
	\begin{definition}[Professional Facilitation]
		\label{def:facilitator}
		Facilitation should occur at the point where a professional facilitator would have intervened. 
	\end{definition}
	
	In current literature, it is generally accepted that all comments made by a facilitator should be considered facilitative interventions; for instance, only these comments are assumed to follow some facilitative strategy~\citep{falk-etal-2021-predicting, chen-etal-2025-whow, schroeder-etal-2024-fora}. Defining facilitative comments this way is simple and intuitive, but can only be used in domains where (1) professional facilitators exist and (2) they can be identified. These conditions exclude the vast majority of online discussions, as very few web-crawled datasets include such information.\footnote{See Table~\ref{tab:datasets}, where the largest datasets such as WikiConv~\citep{wikiconv} do not contain such information.}
	
	\begin{definition}[Human Judgment]
		\label{def:facilitative}
		Facilitation should only occur when observers 
		agree that it is necessary.\footnote{In our study and in \cite{schaffner_community_guidelines}, we mean outside observers, although it could also be observers within the discussion.}
	\end{definition}
	
	There is an argument to be made that professional facilitators can make more accurate judgments due to expertise. However, in reality most facilitation is handled by either volunteers elevated to the position by their community~\cite{schaffner_community_guidelines}, or by the users themselves in an ad-hoc fashion~\cite{umod}. In this case, the task is no longer about detecting or predicting whether \emph{a facilitator would speak}, but rather the much more difficult and subjective task of whether the comment is \emph{facilitative in nature} (as opposed to general participation).
	
	\paragraph{What Is the Difference?} Table~\ref{tab:definitions-example} shows an example where the same discussion is annotated following the three definitions. Def.~\ref{def:escalation} is reliant on the outcome of the discussion, and would miss examples where successful interventions led to a smooth discussion. Def.~\ref{def:facilitator} would give more fine-grained information, although we run the risk of classifying non-facilitative comments as facilitative (and sometimes the opposite~\cite{umod}). Finally, Def.~\ref{def:facilitative} can provide us with fine-grained information, although ambiguous cases remain. It is worth noting that these definitions are not mutually exclusive, as demonstrated in Table~\ref{tab:definitions-example}.
	
	\begin{table}[t]
		\centering
		\small
		\begin{tabularx}{\linewidth}{l X c c c}
			\toprule
			\textbf{Name} & \textbf{Text} & \textbf{D1} & \textbf{D2} & \textbf{D3} \\
			\midrule
			Joe & Thank you for sharing. So I am Joe from Chicago originally... & $\times$ & $\times$ & $\times$ \\
			Marie & Thank you Joe. & $\times$ & \checkmark & ? \\
			Peter & Yeah. So I'll kind of go through it again. I am Peter... & $\times$ & $\times$ & $\times$ \\
			Marie & Thank you, Peter. So I think we have another participant on the line... & $\times$ & \checkmark & \checkmark \\
			\bottomrule
		\end{tabularx}
		\caption{Example text where facilitation interventions are identified according to each definition (D1-3). \textit{Marie} is a facilitator. We assume the discussion does not derail. The second comment \textit{may} be facilitative according to Def.~\ref{def:facilitative}, which depends on the consensus of participants.}
		\label{tab:definitions-example}
	\end{table}

	\subsection{Determining When to Facilitate}
	\label{sec:how_to_determine}
	
	\paragraph{Using Models}
	The only work to our knowledge that has explicitly attempted to predict facilitative interventions~\citep{falk-etal-2021-predicting} uses \ac{eo} classifiers on a single dataset (\acs{ceri}~\citep{ceri}) with modest performance. Indeed, these classifiers have proven competent in the context of escalation prediction~\cite{cmv_awry2,Yuan_Singh_2023}, and facilitative tactics detection~\citep{chen-etal-2025-whow, schroeder-etal-2024-fora}. Recent work~\cite{schroeder-etal-2024-fora, chen-etal-2025-whow} uses LLM-as-a-judge models to annotate data in the field of facilitation, although recent research suggests that fully-automated facilitation is still out of our reach \cite{korre-etal-2025-evaluation}.
	
	\paragraph{Using Humans} 
	Expert human facilitators are frequently used in research~\citep{chen-etal-2025-whow, ceri, schroeder-etal-2024-fora}, but this is not a scalable process due to costs~\citep{Rossi_Harrison_Shklovski_2024}.
	
	\paragraph{Ignoring It}
	Most platforms hold users responsible for maintaining the quality of discussion and the enforcement of community and platform rules, especially in jurisdictions where the platforms themselves are largely absolved of the responsibility of posted content, such as the U.S.~\citep{seering_2020_selfmoderation, schaffner_community_guidelines}. In some cases, the users are responsible for just flagging and reporting content~\citep{seering_2020_selfmoderation}, while platforms like Reddit use volunteer moderators, who also function as community leaders and facilitators~\citep{seering_2020_selfmoderation, umod}, raising ethical questions about ``Moderation as Free Labor''~\citep{Matias2019TheCL}.

	\section{The PEFKI Dataset}
	\label{sec:dataset}
	
	\begin{table*}[t]
		\centering
		\small
		\begin{tabularx}{\textwidth}{Xcccccc}
			\toprule
			\textbf{Name} & \textbf{\#Comments} & \textbf{\#Discussions} & \textbf{Domain} & \textbf{Format} & \textbf{Fac.} & \textbf{Def.} \\ \hline
			WikiDisputes \citeyearpar{wikidisputes} & 96,320 & 8,727 & Wikipedia & Text & $\times$ & \ref{def:escalation}\\ 
			\acf{wt} \citeyearpar{wikitactics} & 3,850 & 213 & Wikipedia & Text & $\checkmark$ & \ref{def:facilitator}\\ 
			WikiConv \citeyearpar{wikiconv} & 17,806,373 & 4,486,983 & Wikipedia & Text & $\times$ & \ref{def:escalation}\\ 
			\acf{awry} \& \acf{cmv} \citeyearpar{cmv_awry2} & 40,607 & 6,691 & Wikipedia \& Reddit & Text & $\times$ & \ref{def:escalation}\\
			\acf{ceri} \citeyearpar{ceri} & 3,700 & 132 & Deliberation & Text & $\checkmark$ & \ref{def:facilitator}\\ 
			\acf{umod} \citeyearpar{umod} & 2,000 & 1,000 & Reddit & Text& $\checkmark$ & \ref{def:facilitative}\\ 
			\acf{iq2} \citeyearpar{zhang-etal-2016-conversational} & 34,245 & 108 & Debate & Oral & $\checkmark$ & \ref{def:facilitator}\\ 
			\acf{whow} \citeyearpar{chen-etal-2025-whow} & 25,542 & 178 & Radio / TV & Oral & $\checkmark$ & \ref{def:facilitator}\\ 
			Fora \citeyearpar{schroeder-etal-2024-fora} & 39,438 & 262 & Story sharing & Oral & $\checkmark$ & \ref{def:facilitator}\\ 
			\bottomrule
		\end{tabularx}
		\caption{The datasets used in \acs{pefk} include a large variety of domains, some crawled, and others as transcripts from oral discussions. All datasets include multi-participant discussions (besides UMOD which contains comment pairs). The \textbf{Fac.} column denotes whether information on which users are facilitators is available and the \textbf{Def.} column which facilitation definition is used in the original dataset (see \S\ref{sec:related:interventions}).}
		\label{tab:datasets}
	\end{table*}
	
	Since there is currently no dataset that explicitly concerns itself with facilitation interventions, we construct the ``\acf{pefk}'' dataset, a standardized, multi-domain compilation of the facilitation-related datasets listed by the work of \citet{korre-etal-2025-evaluation} (see Table~\ref{tab:datasets}). \ac{pefk} contains data from both online forums and oral discussions,\footnote{The differences between different domains and modalities (transcripts vs. written comments) are analyzed in App.~\ref{app:dataset:exploratory}.} and standardizes metadata, metrics, and labeling schemes according to a simple, unified schema, making it suitable as a benchmark on facilitator interventions.\footnote{Information on preprocessing steps and dataset schema can be found in App.~\ref{app:dataset}.} The constituent datasets described below straddle all three definitions described in \S\ref{sec:related:interventions}.\footnote{Including escalation and toxicity, as used in Def.~\ref{def:escalation}, although this information is not utilized in this study.} 
	
	\paragraph{Wikipedia} The WikiConv \citep{wikiconv} dataset is the largest collection of discussions among Wikipedia contributors who may request arbitration by a \emph{volunteer facilitator}. \ac{wt} \citep{wikitactics}, \ac{awry} \citep{cmv_awry2} and WikiDisputes \citep{wikidisputes} are smaller subsets of WikiConv which include additional meta-data information and annotations. The structure of these discussions is ideal for Def.~\ref{def:escalation}, which is why only \ac{wt} includes enough information to be usable for studying facilitation on a deeper level (Def.~\ref{def:facilitator}).
	
	\paragraph{Social Media} The \ac{cmv} \citep{cmv_awry2} and \ac{umod} \citep{umod} datasets are both collected from Reddit, although they answer completely different questions. Like WikiConv, \ac{cmv} looks for discussion escalation by looking for moderation artifacts (here, content moderation in a comment chain) by \emph{volunteer facilitators}. In contrast, \ac{umod} looks for these facilitators via third-party annotation, according to Def.~\ref{def:facilitative}.
	
	\paragraph{Professional Facilitation} In some specialized domains a trained, professional facilitator is present, although their role may vary depending on the specific context of each dataset. In \ac{ceri}, the facilitators were trained by the researchers according to their own developed protocol for deliberation and consensus-building. The same applies for the Fora dataset \citep{schroeder-etal-2024-fora}.\footnote{Although, notably, the level and objectives of this training are not described in the paper.} On the other hand, the facilitators in \ac{whow} and \ac{iq2} are experienced debate hosts (in radio and formal debates respectively).\footnote{Note that some \ac{iq2} debates are also present in \ac{whow}, which is handled in the preprocessing step (App.~\ref{app:dataset:preprocessing}).} These datasets define facilitation according to Def.~\ref{def:facilitator}.

	\section{Do Humans and LLMs Intervene at Similar Circumstances?}
	\label{sec:annotation}
	
	To support a systematic investigation of facilitative behavior under Def.~\ref{def:facilitative}, we conduct a survey on \ac{pefk} to analyze facilitation timing patterns between humans and LLMs. Rather than establishing a ground truth through \ac{iaa}, we adopt a survey-based approach, since prior work has shown that obtaining reliable annotations for similar tasks is highly challenging, particularly in newly explored settings such as ours (see ``Survey Procedure'').

	\subsection{Facilitative Survey}
	\label{sec:annotation:experimental}
	
	\paragraph{Labels} To avoid turning facilitation into a moderation task (i.e., using reactive responses after unwanted comments/utterances), we introduce two types of interventions: positive and negative reinforcement. \emph{Positive reinforcement}, aims to encourage or reward constructive behavior (e.g., expressing appreciation or affirming contributions), while \emph{negative reinforcement}, aims to discourage undesirable behavior or guide participants back to productive discourse (e.g., addressing incivility or off-topic comments). This distinction is just a tool through which we ensured the survey participants did not default to content moderation, which we frequently observed during discussions with them and during pilot survey rounds. It does not represent a theoretical contribution but provides an extra signal which we utilize in our analysis.\footnote{The theoretical grounding for this approach, and its relation to prior work, is described in App.~\ref{app:human-annotation:positive-vs-negative}.} We adopt an ordinal scale approach during annotation, where the survey participants rate each dialogue for either \texttt{Positive}, \texttt{Negative} or \texttt{No} reinforcement using a scale from 1 to 5.\footnote{These are later aggregated to four categories: \texttt{Positive}, \texttt{Negative}, \texttt{Both} and \texttt{None}. See App.~\ref{app:labels} for the exact motivations and mapping of facilitative intervention definitions throughout this study.}
	
	\paragraph{Data}
	We randomly split multi-party discussions from the \ac{pefk} dataset into small chunks containing a set amount of comments (6 comments per discussion), and select a random sample of 1224 discussions for the survey.\footnote{After the participants flagged some excerpts as malformed, 721  remained; see Fig.~\ref{fig:malformed} in App.~\ref{app:human-annotation}.} This design aims to replicate a known limitation in online asynchronous facilitation scenarios; the facilitator is expected to decide on a course of action considering only the last parts of a discussion~\citep{hewitt2003habitual, park2016supporting, kumpel2021does}.\footnote{It also mirrors the conditions of online asynchronous facilitation, where facilitators act on the most recent part of a discussion~\citep{hewitt2003habitual, park2016supporting, kumpel2021does}. A similar pattern appears in oral discussions, where evidence suggests that the most recent comments carry relatively greater weight~\citep{crano1977primacy, cockburn2017effects}.} One important decision was whether to keep or remove facilitative comments. We opted to keep them, preserving the full conversational context, since excluding them might have led to comprehension issues, without any indication of whether a speaker was a facilitator.\footnote{See Limitations for a discussion on this design choice.}
	
	\paragraph{Survey Participants}
	We recruited 11 experts on Computational Social Science with extensive experience with NLP tasks from sentiment analysis to persuasion strategy detection. 
	We preferred expert instead of crowdsourced participants as the task is highly technical, subjective, and mentally demanding. Additionally, crowdsourced participants have been shown to exhibit more bias in general~\citep{snow-etal-2008-cheap, waseem-2016-racist}, which concerned us since we tried to avoid confusion with content moderation (a better known task). Participants gave informed consent in compliance with GDPR~\citep{GDPR} and were compensated in accordance with national legislation. Demographics and selection details can be found in App.~\ref{app:participants_select}.
	
	\paragraph{Survey Procedure} 
	Unlike prior work that sought to establish consensus among facilitators~\cite{umod, schroeder-etal-2024-fora, chen-etal-2025-whow}, our goal is instead to explore facilitation patterns without presuming that such a consensus exists. Indeed, facilitation is a highly subjective task, and substantial disagreements persist even among experienced professionals operating within the same communities and under the same rule-sets~\citep{vinay2024_venire}; consensus remains elusive even in cases where the task is less ambiguous and the dataset more focused~\citep{schroeder-etal-2024-fora}.\footnote{During the survey, we inserted duplicated entries in secret to estimate consistency, finding that most participants remained 100\% consistent in their judgments during the task, see App.~\ref{app:human-annotation:intra-agreement}.}
	
	Before the survey, participants completed a training task categorizing facilitative tactics according to established taxonomies on \ac{pefk}, which we found necessary to disambiguate facilitation from content moderation. To limit potential bias, training covered strategies only, with no input on the timing or flow of conversation (see App.~\ref{app:human-annotation:tactics}).
	
	The main survey task required the participants to rate instances from \ac{pefk} by determining whether, at any point within a specific discussion chunk, reinforcement was needed and what kind. The participants could optionally add rationales with which they could justify their rating. Two pilot surveys, one-on-one interviews and weekly group meetings accompanied the main task (App.~\ref{app:participants_select}). The full survey guidelines can be found in App.~\ref{app:guidelines}.

	\paragraph{LLM Survey}
	We use only open-source models for three reasons. First, proprietary models yield less reproducible results~\citep{bisbee_2023}. Second, although they may perform better, their pricing makes them impractical for large-scale research and real-world deployment. Third, part of our dataset (Fora) is not publicly available and, like the other constituent datasets, may contain personal data. Running local models keeps this data on our own hardware and supports compliance with data protection regulations~\cite{eu_ai_act, GDPR}. We use 4-bit quantized versions from the \texttt{unsloth huggingface} repository. We select models on the basis of prior use in research tasks, model age (younger is better), and availability in the aforementioned repository; specifically \texttt{OLMo 2 \{7B, 32B\}}, \texttt{Qwen 2.5 \{7B, 32B\}} and \texttt{LLaMa 3.1 8B}, \texttt{LLaMa 3.3 70B}. We use zero-shot prompting since no concrete definition of facilitation exists~\citep{korre-etal-2025-evaluation}, and the examples would have to be different for each discussion domain in \ac{pefk}.

	\subsection{Human Facilitation}
	\label{sec:annotation:human}
	
	\begin{figure*}[t]
		\centering
		\includegraphics[width=\linewidth]{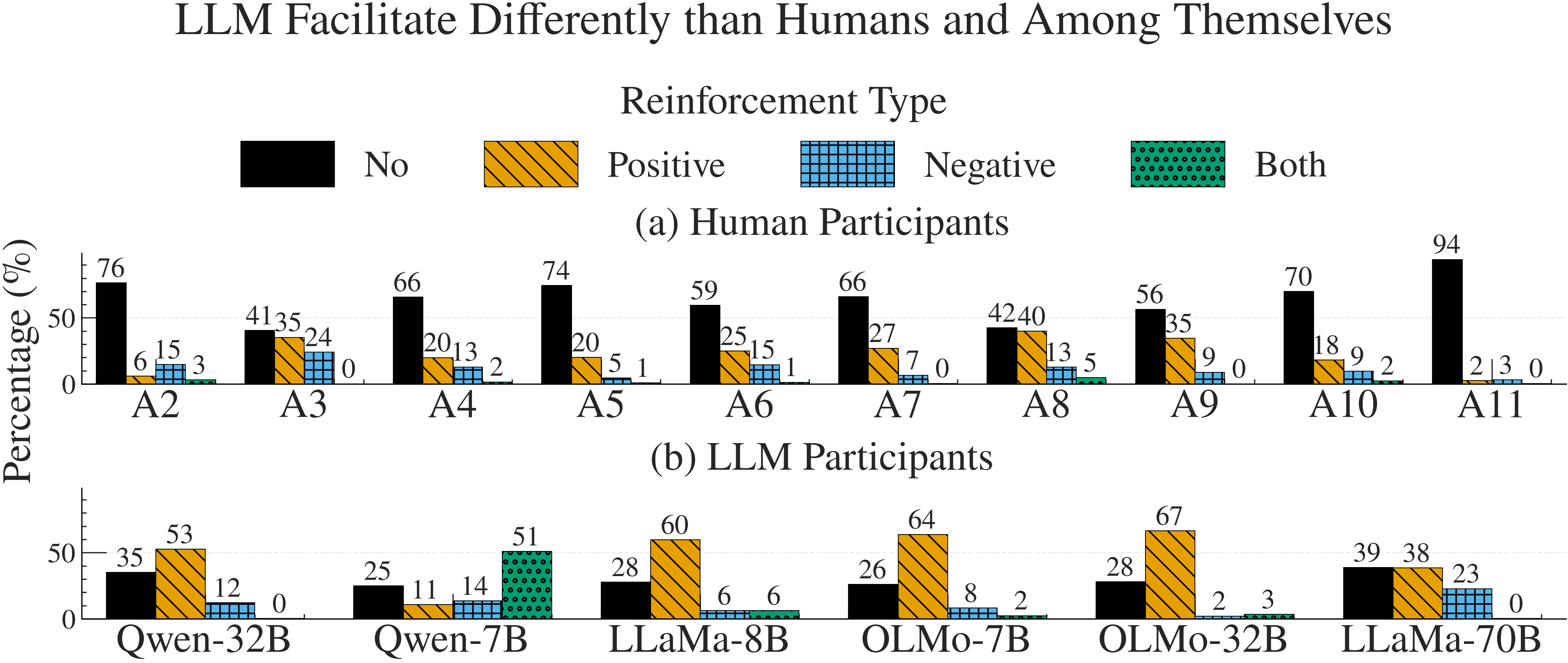}
		\caption{Distribution of all classes per participant. Human participants (a) are either very cautious (A2, A4--A7, A10, A11) or somewhat cautious (A3, A8, A9), unlike LLMs (b) which are significantly more eager to facilitate.}
		\label{fig:freqs_annotators}
	\end{figure*}
	
	\paragraph{Humans Tend to be Cautious Facilitators}
	The main dilemma around facilitative timing is determining whether the facilitator prioritizes preventing potentially harmful speech, or letting the conversation
	continue naturally (Fig.~\ref{fig:intro}).	According to Fig.~\ref{fig:freqs_annotators}, most participants generally remain conservative. Most choose not to intervene in a large majority of cases, although some (e.g., A3, A8, and A9) exhibit comparatively more proactive behavior with higher rates of positive interventions.\footnote{One participant (A1) did not complete the task and was therefore not considered for the analysis.} Positive-intervention rates also appear to cluster around two broad ranges (approximately 20–30\% and 35–40\%), which is consistent with the tendencies of professional facilitators observed in the original dataset (see Fig.~\ref{fig:moderation-perc}; App.~\ref{app:dataset:schema}).\footnote{These observations are consistent across datasets (see App.~\ref{app:human-annotation:datasets}).} 
	
	\begin{figure}[t]
		\centering
		\includegraphics[width=\linewidth]{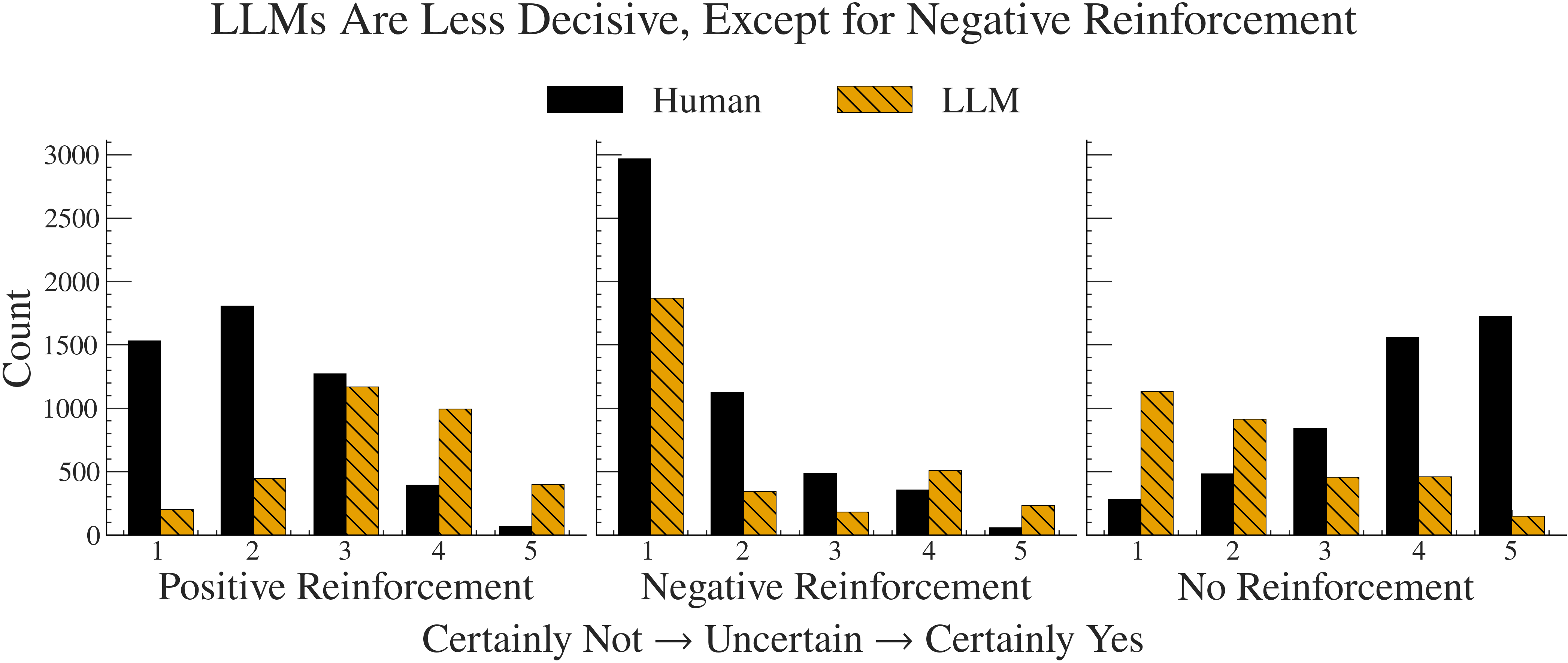}
		\caption{Certainty of judgments per facilitation label. LLMs (orange bars on the right) diverge from humans by favouring positive reinforcement (modes 3--4) where humans mostly disagree (1--2), and reject no reinforcement (1--2) where humans mostly agree (4--5). Both mostly disagree with negative reinforcement (mode 1).}
		\label{fig:freqs_certainty}
	\end{figure}
	
	\paragraph{Human Facilitators May Attend over Different Discussion Spans}
	To qualitatively understand the sources of divergence in human facilitative patterns, we analyzed a ``worst-case'' scenario: situations where half of the participants favored positive reinforcement and the other half favored negative reinforcement. As shown in Table \ref{tab:examples_rationales}, a specific discussion example illustrates this conflict. Participants justifying negative reinforcement cited the conversation's trajectory toward escalation, suggesting intervention was necessary to curb aggressive language. Conversely, the same participants noted that certain comments, particularly the first two, were examples of politeness and hedging, making them candidates for positive reinforcement. 
	
	\begin{table}[t]
		\centering
		\scriptsize
		\begin{tabularx}{\columnwidth}{cX}
			\toprule
			\# & Content  \\
			\hline
			\rowcolor{gray!30}
			1 & Well nothing is really *inherently* bad…  \\
			2 & Well political correctness is about avoiding offense… \\
			\rowcolor{gray!30}
			4 & Are you really arguing that WWI was unjust towards cisgendered men?  \\
			5 & Not WWI itself, but that period, and I don't see what is so odd about it…  \\
			\rowcolor{gray!30}
			6 & There were transgendered people in WWI. You still haven't justified how it was genuinely an injustice… \\
			\bottomrule
		\end{tabularx}
		\caption{Sections from the same discussion chunk. Each is considered an instance, with each row representing a conversation turn. Participants might facilitate differently depending on which section they decide is more important.}
		\label{tab:examples_rationales}
	\end{table}
	
	\begin{figure}[t]
		\centering
		\includegraphics[width=1\linewidth]{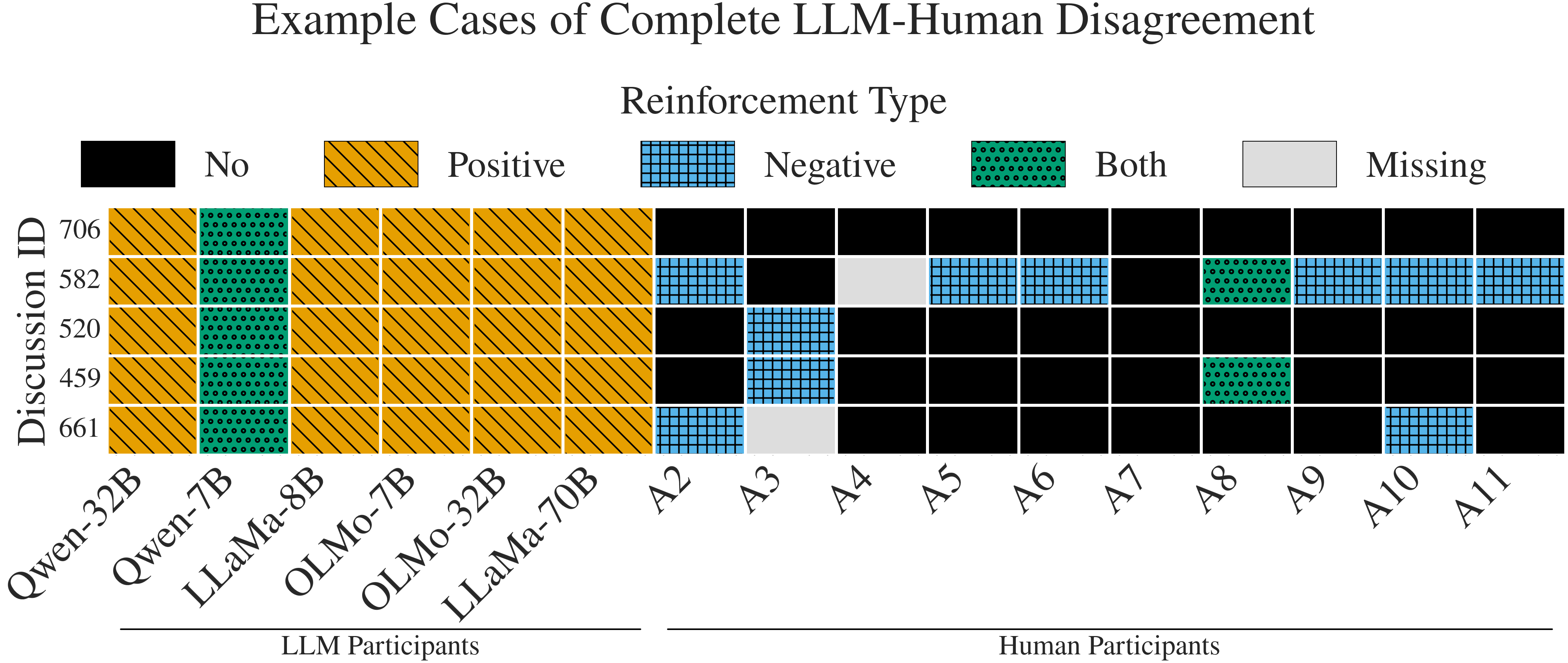}
		\caption{Cases with the highest inter-LLM agreement, where no human agrees with the LLM majority.  Labels follow the survey schema described in App.~\ref{app:human-annotation:positive-vs-negative}.}
		\label{fig:extremes_heatmap}
	\end{figure}
	
	\begin{finding}
		\label{finding:human-facilitate}
		Most human participants tend to be either highly cautious or to balance proactivity with caution. Disagreements may arise because participants focus on different spans of the discussion.
	\end{finding}

	\subsection{Do LLMs Facilitate Differently?}
	
	\paragraph{Yes} Fig.~\ref{fig:freqs_certainty} shows that humans are less certain than LLMs when it comes to positive reinforcement but in both cases the values are concentrated in the middle part of the distribution. For negative reinforcement, both humans and LLMs show high uncertainty, while for no reinforcement there is a clear difference with humans being more certain than LLMs.
	
	\paragraph{Why?} We identify cases where LLMs largely agreed among themselves but completely disagreed with all humans in order to highlight situations where model behavior diverges acutely from human judgment. In total, there are 36 cases where no human agrees with the majority voting of the LLMs. Fig.~\ref{fig:extremes_heatmap} shows the top 5 extreme cases. Despite strong LLM consensus (agreement >= 0.66), all human participants labeled these comments differently. 
	
	Examining the five examples qualitatively, we see that all cases regard a narrative of the participants rather than a debate like discussion that could possibly lead to escalation.\footnote{Unfortunately, we cannot rely on the LLMs for explanations post-hoc, nor use \ac{cot} prompting to extract rationales, since both may produce non-representative and deceptive explanations~\citep{zaman-srivastava-2026-chain, turpin2023language} without extensive evaluation. How the models arrived to these conclusions must therefore be evaluated from our own observations. } Specifically, discussion 706 is about gender, labor market change, and education, where multiple speakers exchange viewpoints on shifting economic roles.\footnote{Since each discussion span can involve as many as 30{,}000 characters, we refer the reader to the supplementary material for context.} The excerpt contains a lot of anecdotal examples (e.g., references to one's workplace examples and ``Aviation High School''), rather than decision-oriented, focusing on broad social interpretation. In general, the dialogue is very self-contained, with speakers building perspectives rather than seeking resolution or guidance. This narrative-like structure may make participants believe that these discussions  comprise ``neutral comments'' (leading to no intervention). LLMs, on the other hand, being more prone to intervention, could interpret this neutrality as an opportunity for positive reinforcement. This is a hypothesis that should be verified with rationales by the participants in future research. An expansion of this analysis on all 36 examples can be found in App.~\ref{app:sec:disagreement_examples}.
	
	\begin{finding}
		\label{finding:llm-facilitate-pefk}
		LLMs are more intrusive and more consistent compared to humans, being eager to view neutral discussions as opportunities for positive reinforcement.
	\end{finding}

	\section{Can Models Predict Interventions?}
	\label{sec:facilitator}
	
	\begin{table}[t]
		\centering
		\small
		\begin{tabular}{lrrrrr}
			\toprule
			Model & Pr. & Rec. & F1$_p$ & F1$_n$ & Sup.\\
			\midrule
			
			\multicolumn{6}{c}{\textbf{\ac{ceri}}} \\
			\midrule
			LLaMa8B  & 0.246 & 0.518 & 0.333 & 0.567 & 335 \\
			ModBert  & 0.319 & 0.490 & \textbf{0.386} & \textbf{0.706} & 192 \\
			{All} & {0.201} & {0.327} & {0.215} & {0.635} & {335} \\
			
			\midrule
			\multicolumn{6}{c}{\textbf{\ac{wt}}} \\
			\midrule
			LLaMa70B & 0.241 & 0.433 & 0.310 & 0.310 & 336 \\
			ModBert  & 0.407 & 0.717 & \textbf{0.520} & \textbf{0.613} & 258 \\
			OLMo32B  & 0.292 & 0.466 & 0.359 & 0.465 & 331 \\
			Qwen32B  & 0.267 & 0.258 & 0.263 & 0.601 & 336 \\
			Qwen7B   & 0.264 & 0.392 & 0.315 & 0.455 & 336 \\
			{All} & {0.307} & {0.581} & {0.393} & {0.361} & {336} \\
			
			\midrule
			\multicolumn{6}{c}{\textbf{Fora}} \\
			\midrule
			LLaMa8B  & 0.386 & 0.584 & 0.465 & 0.562 & 382 \\
			ModBert  & 0.409 & 0.631 & \textbf{0.497} & 0.599 & 2764 \\
			Qwen7B   & 0.412 & 0.102 & 0.164 & \textbf{0.759} & 382 \\
			{All} & {0.390} & {0.355} & {0.279} & {0.604} & {382} \\
			
			\midrule
			\multicolumn{6}{c}{\textbf{\ac{iq2}}} \\
			\midrule
			LLaMa70B & 0.265 & 0.137 & 0.181 & 0.703 & 374 \\
			LLaMa8B  & 0.318 & 0.695 & 0.436 & 0.290 & 374 \\
			ModBert  & 0.399 & 0.634 & \textbf{0.490} & 0.572 & 2393 \\
			Qwen7B   & 0.369 & 0.290 & 0.325 & \textbf{0.693} & 374 \\
			{All} & {0.305} & {0.409} & {0.306} & {0.534} & {374} \\
			
			\midrule
			\multicolumn{6}{c}{\textbf{\ac{whow}}} \\
			\midrule
			LLaMa70B & 0.312 & 0.159 & 0.211 & 0.721 & 364 \\
			LLaMa8B  & 0.349 & 0.778 & 0.482 & 0.343 & 364 \\
			ModBert  & 0.418 & 0.626 & \textbf{0.502} & 0.597 & 1816 \\
			Qwen7B   & 0.370 & 0.294 & 0.327 & \textbf{0.697} & 364 \\
			{All} & {0.364} & {0.428} & {0.326} & {0.561} & {364} \\
			
			\midrule
			\bottomrule
		\end{tabular}
		\caption{Facilitator prediction results for models that did not collapse (F1$_p$ or F1$_n$ fall below 0.15, see Table~\ref{tab:facilitator-prediction-collapsed} in App.~\ref{app:experiments}). Positive-class precision, recall and F1 (F1$_p$), and negative-class F1 (F1$_n$). Best F1$_p$ and F1$_n$ values for each dataset split are highlighted in bold. ``All'' rows report macro averages across all models, including collapsed ones.}
		\label{tab:facilitator-prediction}
	\end{table}

	\subsection{Setup}
	
	\paragraph{Facilitation Definitions} As discussed in \S\ref{sec:related:interventions}, facilitation interventions can be defined in two ways; whether a comment is facilitative in nature (Def.~\ref{def:facilitative}), or whether a dedicated facilitator would participate at a given point in time (Def.~\ref{def:facilitator}).  We frame the task as a classification problem, aiming to identify the last comment made by a non-facilitator before a facilitator intervenes. To this end, we evaluate six open-source LLMs (see previous section) and a ModernBERT classifier \citep{warner-etal-2025-smarter}.
	
	\paragraph{Input} We construct our training and evaluation data from a unified corpus of threaded discussions, where each instance corresponds to a target comment augmented with its conversational context. For every target comment, we retrieve up to $K=3$ preceding comments along the reply chain. Each comment is truncated to a maximum of 3000 characters to control sequence length. Each comment is formatted using XML-style tags\footnote{Context comments are wrapped with \texttt{<CTX>} and the target comment with \texttt{<TGT>}.} with the final sequence containing three context comments in chronological order followed by the target comment with redacted participant names. We use a 60–20–20 split for training the classifier, and choose 2,000 randomly sampled instances evenly distributed across the sub-datasets of \ac{pefk}.
	
	\paragraph{\ac{eo} Models} We fine-tune a pretrained transformer encoder (\texttt{ModernBERT-large}). Training is conducted using the Hugging Face \texttt{Trainer}, weighting the loss function to account for class imbalance and only finetuning the classification head. To improve computational efficiency, we use batches of 64 texts of similar length for 80 epochs and apply early stopping with a warm-up phase. Data loading uses 4 worker processes, and GPU acceleration is utilized when available. We fix the random seed to ensure reproducibility. We use a single Quadro RTX 6000 GPU for the classifier and two GPUs for the LLMs, collectively requiring about 72 hours of computation. 
	
	\paragraph{Ablations} An identical setup is used for an ablation \ac{nlu} task showing that the models can be used to detect whether a comment was made by a professional facilitator to a degree.\footnote{\citet{schroeder-etal-2024-fora} report  facilitation detection scores in the 0.7--0.9 F1 range, which we were not able to reproduce given their \ac{eo} model setup to the best of our ability.} Another ablation study compares training separate \ac{eo} models for oral and written datasets instead of a single model for both, showing that using a single model across all datasets shows reduced performance (App.~\ref{app:experiments:one-model}), and informing our training setup. Finally, we verify that our LLM-survey setup is not influenced by prompt sensitivity via the ablations shown in App.~\ref{app:experiments:ablations}.

	\subsection{Findings}
	
	\paragraph{Deciding When NOT to Intervene Is Easy.}	Table~\ref{tab:facilitator-prediction} shows the results for the facilitator intervention prediction task. We find that while most models are adept at identifying cases where facilitators did not intervene (large $F1_{n}$), they are generally much less successful at identifying when they \emph{would} (poor $F1_{p}$) mirroring the fact that humans are more certain of when not to intervene (\S\ref{sec:annotation:human}).
	
	\begin{finding}
		\label{finding:no-fac-easy}
		Deciding when not to facilitate is easy both for humans and LLMs. Deciding when to facilitate, is not.
	\end{finding}
	
	\begin{figure}[t]
		\centering
		\includegraphics[width=\columnwidth]{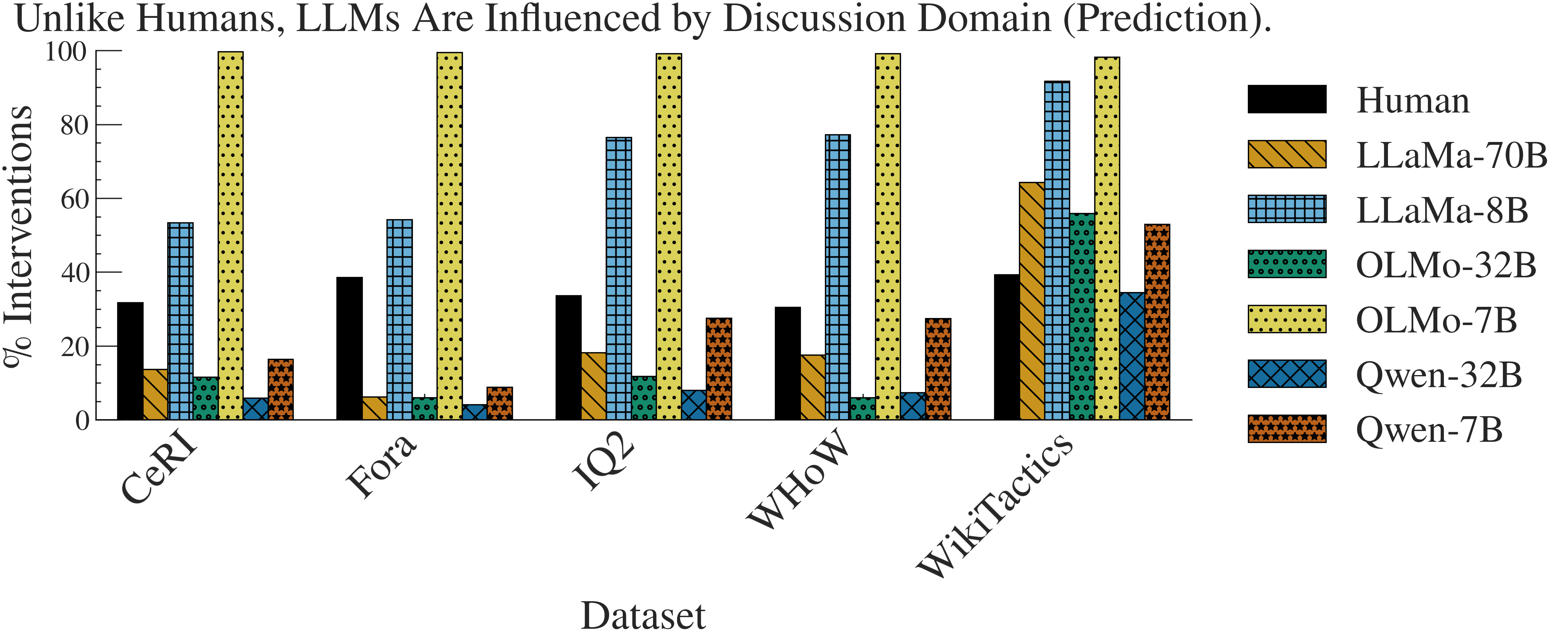}
		\caption{Percentage of times survey participants and LLMs decided to intervene in discussions by dataset. Some LLMs always over-intervene, while others under-intervene in most domains, except for the traditional forum-structured dataset.}
		\label{fig:prediction-by-domain}
	\end{figure}
	
	\paragraph{Do LLMs Over or Under-Intervene?}  LLM behavior in facilitation is polarized: some models exhibit a pervasive tendency to constantly facilitate (reflected by extremely low $F1_n$). Conversely, we observe that other models rarely intervene (evidenced by extremely low $F1_p$, both crossed out in Table~\ref{tab:facilitator-prediction}), a behavior that has been largely overlooked in the existing literature. Interestingly, this pattern also varies by domain; most LLMs tend to under-facilitate in non-forum domains, and over-facilitate in \ac{wt}, the only dataset featuring a traditional forum structure, as shown in Fig.~\ref{fig:prediction-by-domain}. We hypothesize that  this discrepancy is caused by \ac{wt} being the only dataset considered in this study specifically containing escalating discussions, which likely triggers LLM interventions. Human facilitators do not seem to be influenced by the escalating discussions however (Fig.~\ref{fig:prediction-by-domain}), indicating that LLMs may be hyper-sensitive to unsafe speech because of their extensive alignment procedures.
	
	\paragraph{Are \ac{eo} Models the Solution?} We find that the ModernBERT classifiers can generally compete with LLMs in this task, and seem much more consistent, outperforming them in three out of the five datasets, and scoring the highest $F1_p$ scores across all datasets. We posit that overall, these classifiers may be a much more scalable, reliable and viable solution compared to LLMs for the task, even though they have been overlooked in current literature.\footnote{Another notable advantage of these models is that we can tune the precision-recall trade-off by adjusting the probability threshold over which the model determines that an intervention should happen, thus configuring how proactive or cautious they are, see App.~\ref{app:experiments:thresholds}.}
	
	\begin{finding}
		\label{finding:llms-may-facilitate}
		Off-the-shelf LLMs are generally not consistent in deciding when to facilitate, in contrast to traditional \ac{eo} classifiers.
	\end{finding}

	\section{Conclusion}
	We investigated how the timing of humans and LLMs differs in facilitative interventions. By conducting a literature review, we isolated three main definitions of facilitative interventions. We then conducted a survey task using facilitative social science experts on a standardized, aggregated corpus of all relevant facilitation datasets, finding that human facilitators tend to adopt a cautious approach, are more confident when deciding not to intervene than when deciding to intervene, and may disagree because they focus on different spans of the discussion. We are the first to conduct extensive experiments on facilitative intervention prediction using LLMs, finding that they tend to over-intervene, skew towards positive reinforcement when the situation is ambiguous, and remain inconsistent across facilitation decisions, while simpler ModernBERT classifiers trained on existing datasets can outperform them.
	
	Despite optimism that LLMs would revolutionize facilitation, our findings suggest that progress may be incremental, not transformative. LLMs cannot meaningfully facilitate until we move beyond limited datasets. Previous approaches, such as focusing on escalation or general moderation, provided large, scalable data, but only captured surface-level behaviors. They failed to give us a true understanding of how human facilitators operate, a complex and open question that cannot be answered without annotated data that captures genuine facilitative dynamics.

	\section*{Limitations}
	
	\paragraph{Scope and Scale}
	The participants saw discussion excerpts rather than full threads, which can be impractically long. However, real-world facilitation decisions also depend on long-term dynamics, participant history, and community norms that excerpts may not capture. Domain-specific facilitative norms may also have affected LLM performance, but these norms are unexplored in the literature, and prompt optimization is beyond our scope. It is also worth noting that \ac{pefk} is predominantly English and drawn from specific online and institutional settings, so our findings may not generalize to multilingual or culturally distinct contexts.
	
	\paragraph{Data Collection}
	Because providing rationales was optional for participants, we collected very few. This information is valuable and should be incorporated in future tasks concerning facilitation. We only realized their usefulness during the analysis of our results, although it has to be acknowledged that making them mandatory would have significantly added to the workload of the participants.
	
	\paragraph{Intervention Categorization}
	Ideally, the task should be conducted without the use of different intervention labels or very fine-grained labels, as prior work has shown that different domains call for different facilitative tactic taxonomies~\citep{ceri, chen-etal-2025-whow, schroeder-etal-2024-fora}. The decision to include the distinction between negative and positive reinforcement was motivated by the fact that the participants were over-focusing on the content moderation part during the refinement stage of the task. 
	
	\paragraph{Annotation Bias}
	We acknowledge that our choice of leaving (unmarked) facilitative comments in the annotation procedure may have biased annotations, as they could have reacted to the facilitator's interventions rather than to the exchange between participants. We deemed this a necessary risk, since removing these comments made some discussions incomprehensible. It would also alert the annotators that a facilitator comment was removed, which would bias the annotation anyway.

	\section*{Ethical Considerations}
	
	\paragraph{Annotation} All experiments and data received official IRB approval by the \textit{``Athena'' Research Center Ethics Committee (EHDE)} (Protocol Authorization Number: \texttt{66-17/11/2025}). The participants were selected via formal recruitment methods based on familiarity with the task and field, and were compensated fairly with a monthly compensation of 1000€.\footnote{Ministry of Development, Hellenic Republic, Public Call \texttt{GD.439.ARCHI-YP-0625}. See the published document via this ID for details.} They signed informed consent forms before beginning the survey (App.~\ref{app:consent}), were informed of and consented to potentially offensive or inappropriate content, and were allowed to skip uncomfortable tasks at any point, with no required justification (see Ethics Committee protocol). All necessary data were anonymized both during data collection and in the artifacts presented in this work, while non-essential information was destroyed following the conclusion of the study in accordance with GDPR~\citep{GDPR} and the EU AI Act~\citep{eu_ai_act}. Participant demographic data was not collected besides the aggregates shown in this study.
	
	\paragraph{Reproducibility} We use exclusively open-source models, libraries and LLMs, to ensure reproducibility, protect user data, and comply with relevant regulations. The experiment code is publicly released under the \ac{gpl} open-source license. We release other artifacts under a CC-BY-SA-4.0 International License.
	
	\paragraph{Licensing} To help research in the field of facilitation, we openly release the code that constructs \ac{pefk} under a \ac{gpl} license. All datasets used in \ac{pefk} are obtained from the sources provided by the respective authors and are consistent with intended use (academic research). We do not distribute the corpus separately, and thus no additional data license is applicable. The licenses of the original distribution apply. The Fora dataset \citep{schroeder-etal-2024-fora} is not publicly available because it may contain personally identifying information; researchers should contact the authors of Fora for access.
	
	\paragraph{Dual Use} Models that predict when to intervene in discussions could be misused to covertly steer discourse or help automated accounts mimic facilitators. The facilitator-detection experiments could likewise be used to identify or imitate facilitators. We consider the immediate risk limited, as our models perform modestly and inconsistently. Additionally, our work can be used to check when this assumption no longer holds given the release of future, stronger models. We recommend against unsupervised deployment and in favor of human oversight and disclosure to participants.\footnote{Such requirements already exist in some regulatory frameworks, such as the EU AI Act \citep{eu_ai_act}.} Understanding how LLMs behave as facilitators may also help in detecting and auditing their misuse.
	
	\section*{Acknowledgments}
	
	\paragraph{Funding} This work has been partially supported by project MIS 5154714 of the National Recovery and Resilience Plan Greece 2.0 funded by the European Union under the NextGenerationEU Program.
	
	\paragraph{AI Use Statement} We used two Gemma-4 \{5B, 12B\} models (\texttt{gemma-4-E4B-it-Q4\_K\_M.gguf}, \texttt{gemma-4-12b-it-Q4\_0.gguf}) locally-hosted using the \texttt{textgen} software (\url{https://github.com/oobabooga/textgen}) and Anthropic's Sonnet 5.5 model for orthographic and grammatical support. The Gemma4-12B model was also used to summarize the Consent Form (App.~\ref{app:consent}), while Sonnet 5.5 was used for code generation (specifically graph creation, \ac{eo} model setup, and general code refactoring and organization). All outputs have been verified by the authors.
	
	\paragraph{Contributions} We outline the contributions of each member of the research team below, according to the standardized CRediT taxonomy \citep{credit}.
	% CRediT
	\credit{DT}{Conceptualization, Data curation, Methodology, Formal analysis, Investigation, Software, Visualization, Writing -- original draft, Writing -- review \& editing, Resources}
	\credit{KK}{Conceptualization, Data curation, Methodology, Formal analysis, Investigation, Validation, Visualization, 	Writing -- original draft, Writing -- review \& editing}
	\credit{JP}{Conceptualization, Funding acquisition, Methodology, Project administration, Supervision, Writing -- review \& editing}
	
	\insertcredits
	
	\bibliography{refs}
	\appendix
	
	\printacronyms[heading=section]

	\section{Dataset Details}
	\label{app:dataset}
	
	\begin{table*}[t]
		\centering
		\begin{tabular}{llp{10cm}}
			\toprule
			\textbf{Name} & \textbf{Type} & \textbf{Description} \\
			\midrule
			conv\_id & string & The discussion's ID. Comments under the same discussion refer to the same discussion ID. \\
			message\_id & string & The message's (comment's) unique ID. \\
			reply\_to & string & The ID of the comment which the current comment responds to. nan if the comment does not respond to another comment (e.g., it's the Original Post (OP)). \\
			user & string & Username or hash of the user that posted the comment \\
			is\_moderator & bool & Whether the user is a facilitator. \\
			moderation\_supported & bool & True if the moderation labels are directly computed from the original dataset \\
			escalated & bool & A discussion-level measure denoting discussions which have been derailed \\
			escalation\_supported & bool & True if the escalation labels are directly computed from the original dataset \\
			text & string & The contents of the comment \\
			dataset & string & The dataset from which this comment originated from \\
			notes & JSON & A dictionary holding notable dataset-specific information \\
			toxicity & float & The ``toxicity'' score given to the comment by the \citet{google_perspective_api} \\
			severe\_toxicity & float & The ``severe toxicity'' score given to the comment by the \citet{google_perspective_api} \\
			\bottomrule
		\end{tabular}
		\caption{The full set of columns used for each comment in the \ac{pefk} dataset.}
		\label{tab:dataset-columns}
	\end{table*}

	\subsection{Preprocessing}
	\label{app:dataset:preprocessing}
	
	\paragraph{General}
	\begin{itemize}
		\item We exclude comments with no text.
		\item We exclude discussions with fewer than two distinct participants.
		\item We exclude discussions which are common between wikitactics and wikiconv as well as wikidisputes and wikiconv.
		\begin{itemize}
			\item There may be duplicate discussions between wikidisputes and wikitactics, but we allow them since they feature complementary information.
		\end{itemize}
	\end{itemize}
	
	\paragraph{Wikiconv}
	\begin{itemize}
		\item The Wikiconv corpus does not contain information about which user is a moderator/facilitator. Therefore, all comments relating to Wikiconv are tagged as non-moderators.
		\item Additionally, we follow the instructions of the original researchers \cite{wikiconv}, and select only discussions which have at least two comments by different users.
		\begin{itemize}
			\item Wikipedia (thankfully) does not track users who log in with only an IP address (in the original dataset, their \texttt{user\_id} is always set to 0 and their username is of the form \texttt{211.111.111.xxx}). We consider each such username to be a separate user.
			\item Due to the size of the dataset, we have to partially load it during preprocessing. Thus, there is a small chance every 100,000 records that a discussion is marked as a false negative and a part of it gets discarded.
			\item We only include English comments in the final dataset. We use a small, efficient library (\texttt{py3langid}) for language recognition, due to the large size of Wikiconv. Non-English comments are discarded before selecting valid discussions (see point above).
		\end{itemize}
	\end{itemize}
	
	\paragraph{Wikitactics}
	\begin{itemize}
		\item We infer facilitative actions by whether the comment belongs in any of the following categories:
		\begin{itemize}
			\item Asking questions
			\item Coordinating edits
			\item Providing clarification
			\item Suggesting a compromise
			\item Contextualisation
		\end{itemize}
		\item The above tactics are a subset of the Coordinative labels used in the WikiTactics paper. They were selected because they are not used necessarily on 1-1 discussions; they could reasonably be applied by third-party participants. Contrast them with other Coordinative labels such as ``Conceding/recanting'' and ``I don't know''.
	\end{itemize}
	
	\paragraph{Wikidisputes}
	\begin{itemize}
		\item Since only 0.03\% of the comments in the dataset are made by moderators, we mark the dataset as not supporting moderation.
	\end{itemize}
	
	\paragraph{UMOD}
	\begin{itemize}
		\item Facilitative actions are marked as a gradient from 0 (no facilitation) to 1 (full facilitation). We adopt a threshold of 0.75 to consider an action as facilitative, with more than 50\% participant agreement (measured as entropy in the original dataset).
	\end{itemize}
	
	\paragraph{CMV-AWRY2}
	\begin{itemize}
		\item We mark a discussion as escalated when the derailment value (from the official dataset) is in the 60th upper percentile.
		\item We remove deleted (``[deleted]'') comments.
	\end{itemize}

	\subsection{Schema}
	\label{app:dataset:schema}
	
	\begin{figure}[t]
		\includegraphics[width=\columnwidth]{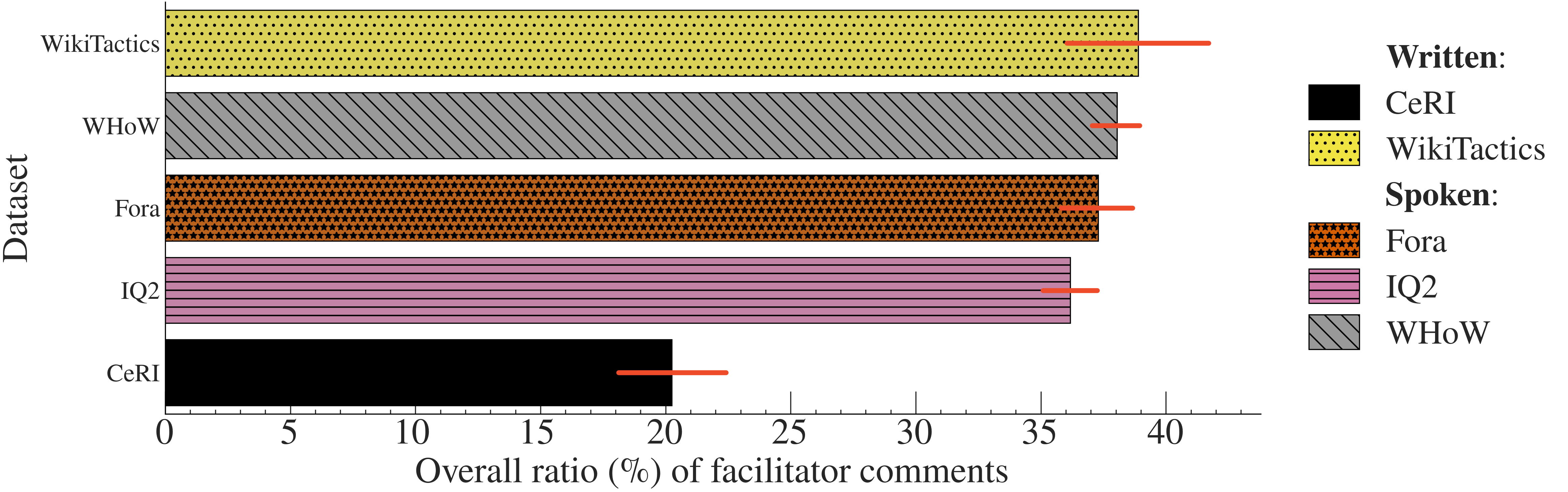}
		\caption{Percentage of comments made by professional facilitators (see Def.~\ref{def:facilitator}) for each of the sub-datasets of \ac{pefk}. Bars denote 95\% confidence intervals. Despite the different kinds of facilitators in online platforms and various real-life-discussion domains, the overall rate of facilitation remains similar across datasets.}
		\label{fig:moderation-perc}
	\end{figure}
	
	Table~\ref{tab:dataset-columns} details the information available for each comment in the dataset. Some information, such as whether the person talking is a moderator / facilitator, is available only on some sub-datasets. Also note that the \texttt{is\_moderator} column may hold different definitions of what we consider  as moderator / facilitator. \ac{umod} for instance considers facilitative comments made by users, while others such as Fora consider a comment facilitative if it was made by a dedicated facilitator.

	\subsection{Exploratory Data Analysis}
	\label{app:dataset:exploratory}
	
	\begin{figure*}[!t]
		\includegraphics[width=\columnwidth]{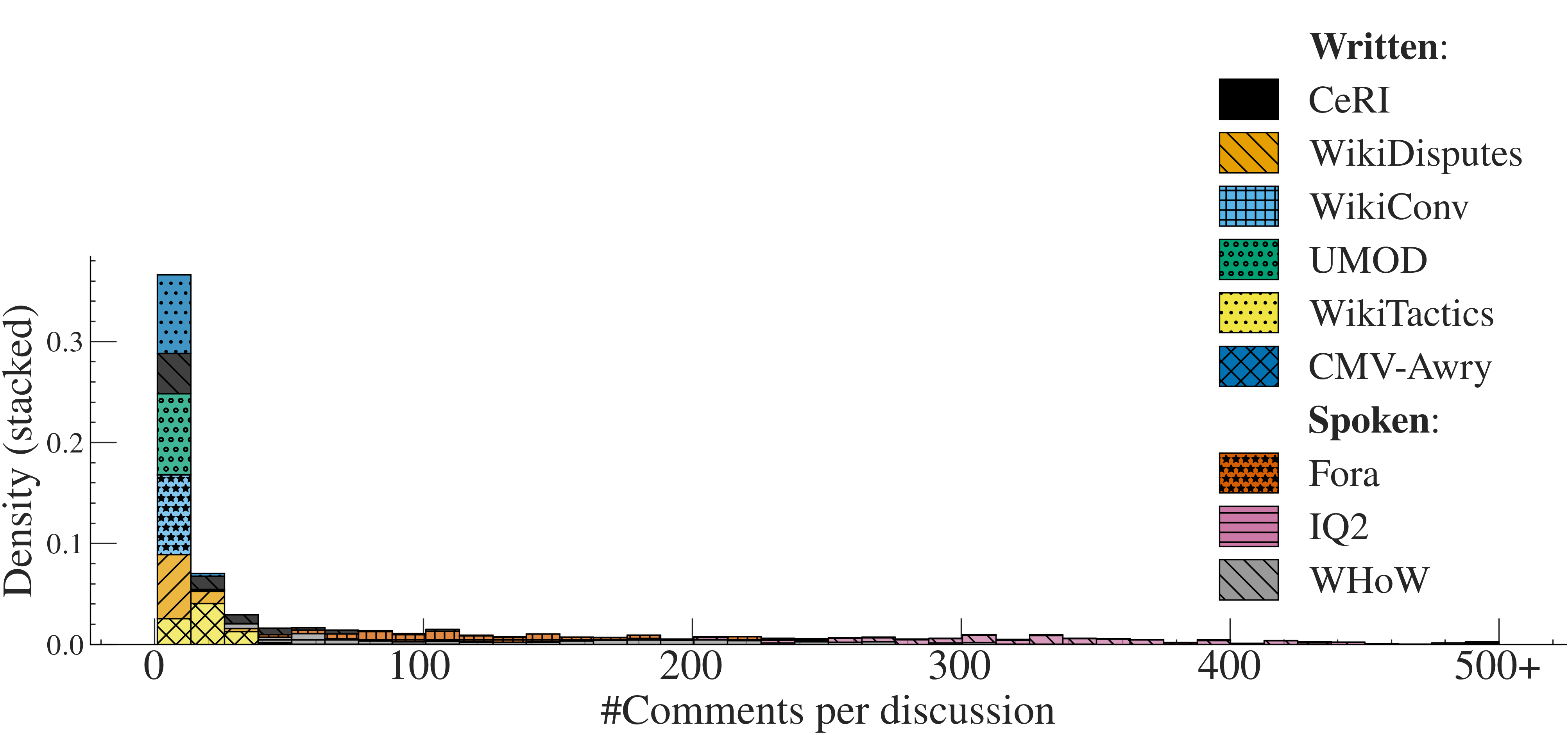}
		\includegraphics[width=\columnwidth]{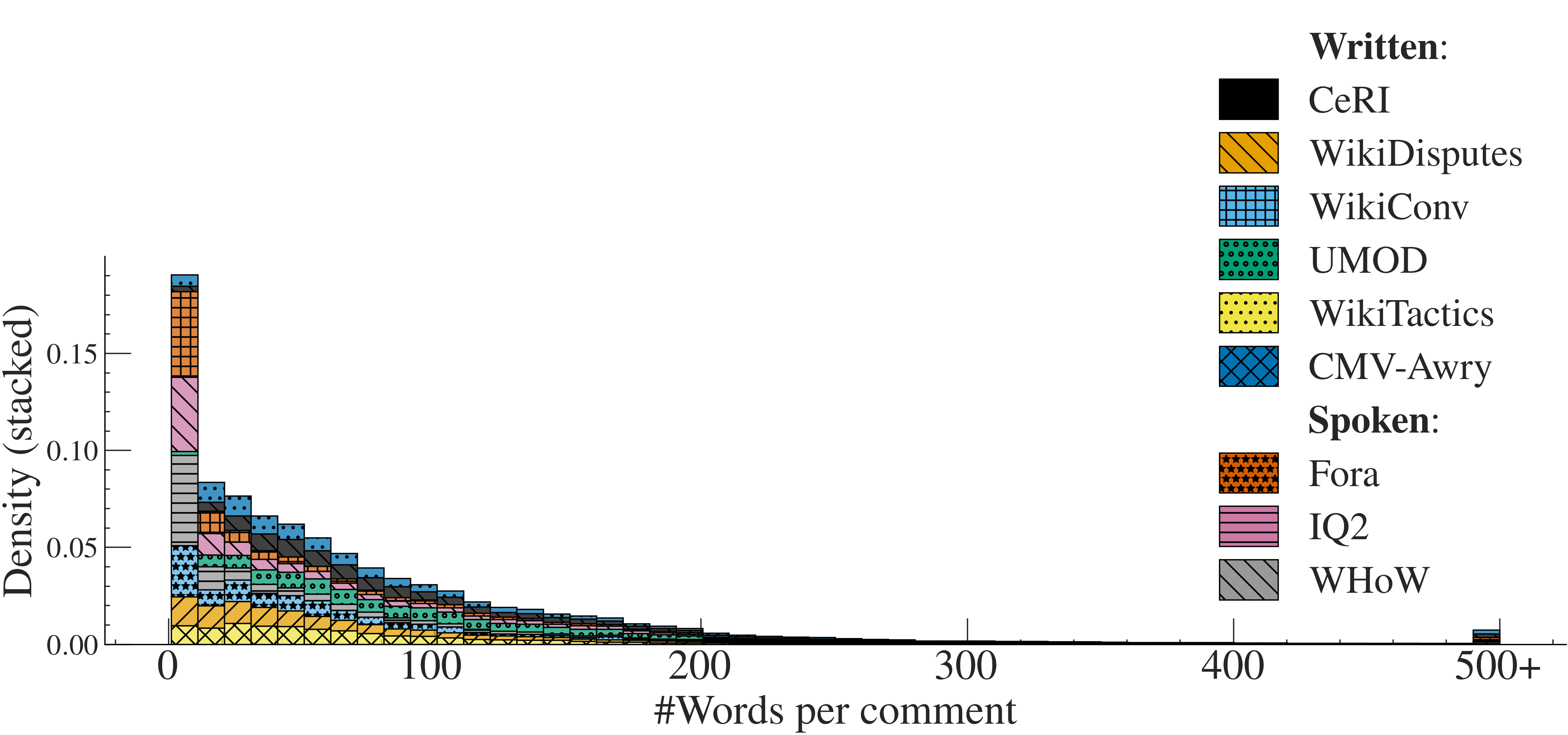}
		\caption{\textbf{Left:} Comments per discussion on the \ac{pefk} dataset. Note the significant increase in comments made in transcribed datasets such as Fora and IQ2. This deviation is explained by the fact that a single human utterance is regularly broken up to many comments in order to simulate interruptions or participants talking over each other. \textbf{Right:} Number of words per comment in all discussions in \ac{pefk}, grouped by original dataset. Comments across written datasets follow the same distribution, while oral datasets feature a significant number of comments with very few words.}
		\label{fig:dataset-stats}
	\end{figure*}
	
	\paragraph{Oral vs. Written Datasets in PEFKI}	
	There currently exists a rift in facilitation research; while some previous work concerns itself with forums and online discussions~\citep{ceri}, others are exclusively focused on live, real-life, oral discussions~\citep{schroeder-etal-2024-fora, chen-etal-2025-whow}. This rift is an issue for facilitation research, since oral and online written discussions feature fundamentally different kinds of facilitation interventions, as shown by stark differences in proposed facilitation schemas between online forums~\citep{ceri}, and real-life discussions~\citep{schroeder-etal-2024-fora, chen-etal-2025-whow}. Additionally, the data themselves are different; posts and comments are self-contained, while transcripts regularly feature interruptions, participants talking in parallel (see App.~\ref{app:dataset:exploratory}), and linguistic features that are necessary for coordination (e.g., ``Thank you.'', ``I agree.''). It is worth noting that in our case, all written discussions are asynchronous (the participants may join and leave at any time), while all oral discussions are synchronous (the participants are all active at the same time).
	
	Figure~\ref{fig:dataset-stats} shows some of the differences of oral and written datasets. Oral datasets feature a significantly larger number of comments per discussion, since instances where the participants talk over each other, or interrupt the other are transcribed as separate comments. Additionally, oral datasets feature a large amount of comments with very few words, since a lot of utterances are short but necessary for social co-ordination (e.g., ``Thank you'', ``Yes'', ``I agree''). Figure~\ref{fig:moderation-perc} shows the percentage of comments made by facilitators for each of our datasets. In almost all datasets, a third of the comments are made by facilitators.\footnote{We do not include the \ac{umod} dataset, since there is no established ground-truth (see App.~\ref{app:dataset:preprocessing}).}
	
\begin{figure}[t]
		\centering
		\includegraphics[width=\linewidth]{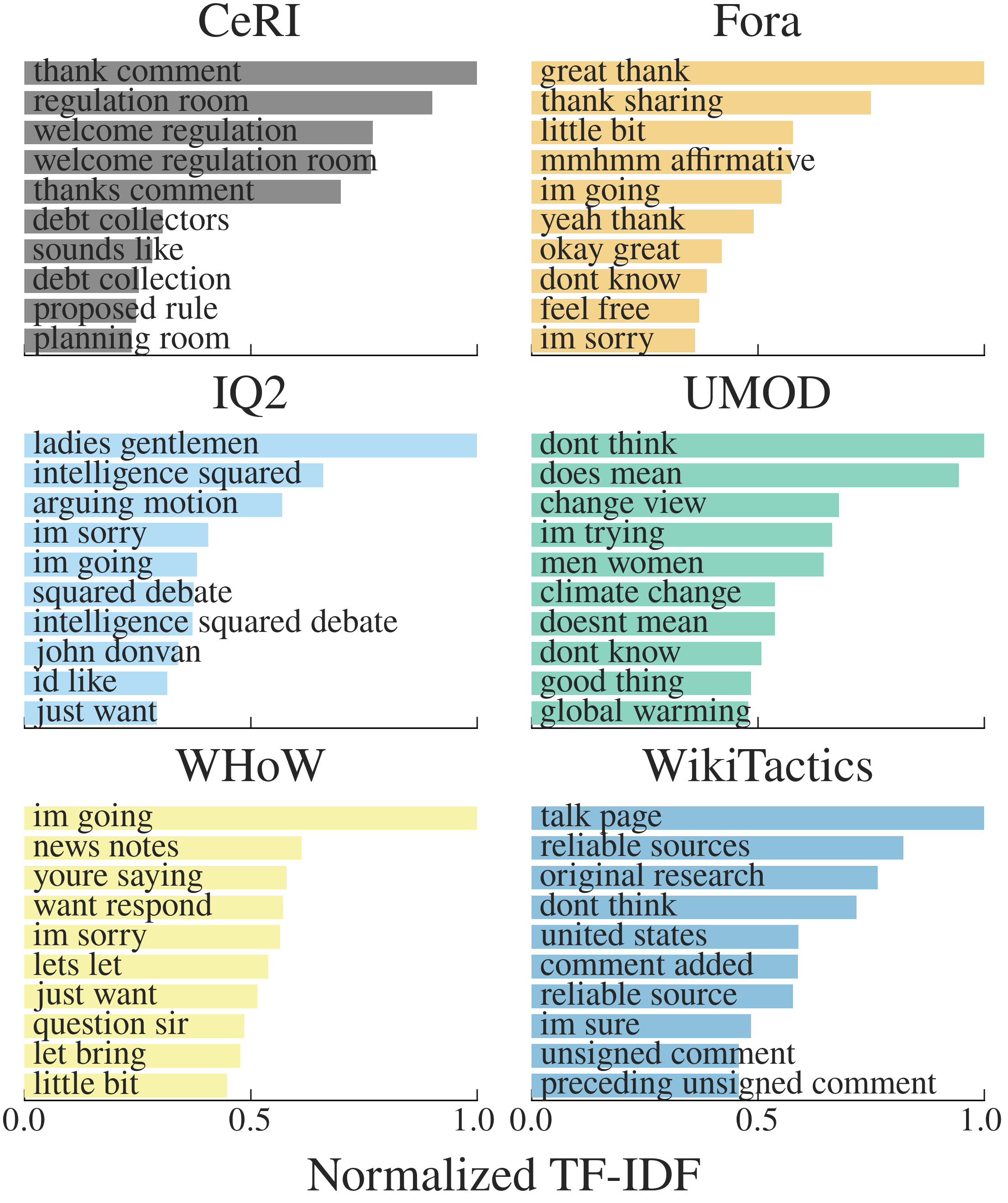}
		\caption{Most common moderator bigrams in \ac{pefk} per dataset. Function words are excluded.}
		\label{fig:pefk_words}
	\end{figure}

	\paragraph{Vocabulary discrepancies in PEFKI}
	After preprocessing, we used a TF-IDF vectorization approach with n-grams (bigrams and trigrams) to extract phrases. Crucially, the TF-IDF model was trained on all datasets combined, so that common, generic phrases were downweighted while phrases that are more distinctive to each dataset were emphasized. After restricting to moderator comments and preprocessing, we computed the average TF-IDF score of each phrase within each dataset and selected the top-ranking ones as representative expressions of moderator language in that context. The results can be found in Figure~\ref{fig:pefk_words}.
	
	The most important phrases in \ac{ceri} suggest a formal, policy-oriented environment with an emphasis on structured participation and regulatory discussion (e.g., references to ``proposed rule'' and ``welcome regulations''). In \textit{fora}, the language is more conversational and facilitative, with phrases such as ``great thank'', ``feel free'', and ``awesome thank'' indicating a more informal moderation style. The \ac{iq2} dataset reflects the debate setting, with phrases like ``ladies gentlemen'' and ``arguing motion''. In \ac{umod}, moderators engage more in argumentation and clarification, as seen in phrases like ``change view'', ``missing point''  reminiscent of more facilitative settings and suggesting a focus on reasoning and discussion quality. The \ac{whow} dataset appears similar to moderated discussions or panels, with phrases like ``question sir'' and ``answer question'' indicating turn-taking and audience interaction. Finally, \ac{wt} shows a highly procedural and rule-based style, with phrases such as ``reliable sources'', ``original research'' and ``talk page'' reflecting the norms and policies of collaborative editing environments like Wikipedia.

	\section{Facilitative Labels Through the Study}
	\label{app:labels}
	
	\begin{figure}[t]
		\includegraphics[width=\columnwidth]{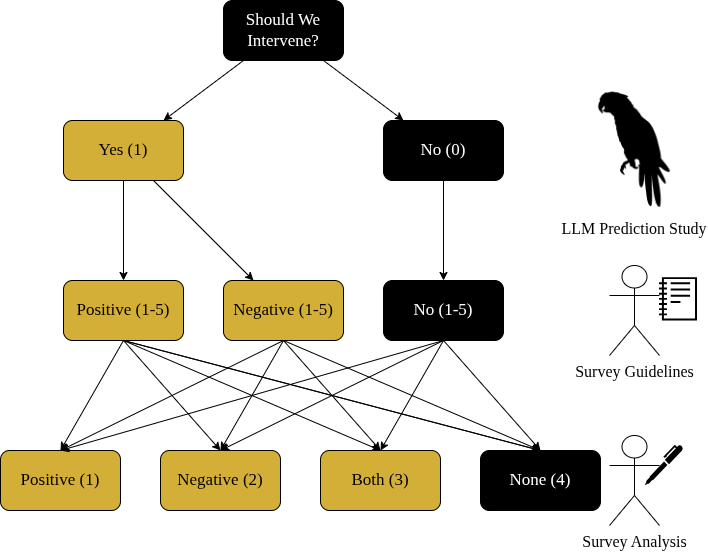}
		\caption{Facilitative labels throughout the different parts of this study. Parrot image taken from \url{https://svgsilh.com/image/1636483.html}, released under CC0.}
		\label{fig:labels}
	\end{figure}
	
	\begin{table}[t]
		\centering
		\begin{tabular}{cccc}
			\toprule
			\textbf{Positive} & \textbf{Negative} & \textbf{Neither} & \textbf{Final} \\
			\midrule
			\xmark & \xmark & \xmark & None (0) \\
			\cmark & \xmark & \xmark & Positive (1) \\
			\xmark & \cmark & \xmark & Negative (2) \\
			\xmark & \xmark & \cmark & None (0) \\
			\cmark & \cmark & \xmark & Both (3) \\
			\cmark & \xmark & \cmark & Positive (1) \\
			\xmark & \cmark & \cmark & Negative (2) \\
			\cmark & \cmark & \cmark & Both (3) \\
			\bottomrule
		\end{tabular}
		\caption{Reinforcement label transformations.}
		\label{tab:label_transform}
	\end{table}
	
	While the main research question of this paper (``When do we facilitate'') is a clear binary decision (we either intervene, or we do not), in practical terms, we had to analyze this question in different ways throughout this study. Fig.~\ref{fig:labels} shows which definitions have been used for each step of this study.
	
	\paragraph{Binary Labels} The binary decision is well-defined for the facilitative prediction experiments shown in \S\ref{sec:facilitator} (as well as those of facilitative detection in App.~\ref{app:experiments:facilitator}), as the model needs to make a single decision based on Def.~\ref{def:facilitator}. 
	
	\paragraph{Positive vs. Negative} On the other hand, \S\ref{sec:annotation} uses Def.~\ref{def:facilitative}, and as such requires a solid understanding of the goals of facilitative interventions. Because of our participant's bias towards negative interventions (as discussed in \S\ref{sec:annotation}), we expanded the binary definition of facilitation to three categorical variables: \texttt{Positive}, \texttt{Negative} and \texttt{None}. Furthermore, since the task is inherently subjective, each of these categories was given a ordinal confidence rating of 1--5 in each response. 
	
	\paragraph{Positive AND Negative} Finally, during analysis we realized that in many cases the survey participants decided that facilitation should occur, but could not discern whether this intervention should be positive or negative. In terms of our previous categorical schema, this phenomenon manifested as low ratings for \texttt{None} and simultaneously high ratings for \texttt{Positive} and \texttt{Negative}. In order to not lose this potentially valuable information, we transform the fine-grained, ordinal label per type adopting a multiclass schema (\texttt{Positive}, \texttt{Negative}, \texttt{None}, \texttt{Both}), as shown in Table~\ref{tab:label_transform}.

	\section{Facilitative Prediction Experiments}
	\label{app:experiments}
	
	\subsection{Can LLMs understand What Facilitation Is?}
	\label{app:experiments:facilitator}
	
	One way of establishing whether the models can understand the concept of facilitation is exploring whether these models can distinguish between facilitative and non-facilitative comments in the first place. Their performance in this task should give us an indication of how well-defined the task is in relevant \ac{nlp} datasets. We thus explore \emph{``facilitator detection''}, an \ac{nlu} task which aims to identify whether a comment was made by a facilitator (Def.~\ref{def:facilitator}). This task has been explored exactly once \citep{shiota2018_facilitators}, although the authors used hand-crafted features instead of relying exclusively on the language used by the facilitators, and used only role-play focused scenarios instead of real-life facilitation ones.
	
	Table \ref{tab:facilitator-detection} presents the performance of these same models in the task of facilitation detection. While we observe instances of strong performance (such as Qwen32B achieving high scores on \ac{ceri} (F1$_p$=0.904, F1$_n$=0.971)) the models generally do not score reliably across diverse datasets (e.g., LLaMa-70B in \ac{wt} yields F1$_p$=0.500). This suggests that the fundamental cause of under-performance is less a confusion with the traditional moderator role and more a structural issue with the datasets themselves. Prior work utilizing traditional \ac{ml} \citep{shiota2018_facilitators}, LLMs, and \ac{eo} classifiers shows similar challenges on other tasks \citep{chen-etal-2025-whow}, indicating that this finding is not merely a result of our specific labeling scheme or dataset selection. Thus, significant, generalizable improvements can only be realized by incorporating larger professional facilitation datasets, where interventions can be \emph{reliably defined and distinguished} from non-facilitative comments, or by developing larger-scale corpora that contain annotated \emph{facilitative comments} (Def.~\ref{def:facilitative}), an extremely difficult task given its inherent subjectivity (\S\ref{sec:annotation}).

	\subsection{LLM Sensitivity Ablations}
	\label{app:experiments:ablations}
	
	We run an ablation study using five instruction prompt variants on a 20\% sample of the dataset. We generated these variants via a local Gemma-4 model (specifically, the \url{https://huggingface.co/unsloth/gemma-4-E4B-it-GGUF} variant, generated using the TextGen open-source software\footnote{\url{https://github.com/oobabooga/textgen}).} in order to avoid researcher bias, and verified the resulting output. Two of these variants include specific instructions warning against over-intervention (``cautious 1,2''). The pairwise agreement scores for the six models can be found in Table~\ref{tab:pairwise_similarity} and the instruction variants in App.~\ref{app:guidelines:llm:ablation}. We find that LLM judgments seem to remain broadly consistent, even with large changes to the LLM input for the specific task. Some inconsistencies do exist, as is the case with all LLM-related tasks~\citep{bisbee_2023}, and some models (such as Qwen-7B) may be more sensitive than others.
	
	\begin{table*}[t]
		\centering
		\small
		\begin{tabular}{llcccccc}
			\toprule
			\textbf{Run A} & \textbf{Run B} & \textbf{LLaMa-70B} & \textbf{Qwen-32B} & \textbf{OLMo-32B} & \textbf{OLMo-7B} & \textbf{Qwen-7B} & \textbf{LLaMa-8B} \\
			\midrule
			cautious1 & cautious2   & 0.975 & 0.978 & 0.888 & 0.972 & 0.966 & 0.972 \\
			cautious1 & concise     & 0.880 & 0.771 & 0.771 & 0.808 & 0.756 & 0.776 \\
			cautious1 & descriptive & 0.759 & 0.870 & 0.739 & 0.760 & 0.587 & 0.741 \\
			cautious1 & role-play   & 0.945 & 0.664 & 0.799 & 0.689 & 0.775 & 0.807 \\
			cautious2 & concise     & 0.867 & 0.784 & 0.898 & 0.807 & 0.760 & 0.767 \\
			cautious2 & descriptive & 0.785 & 0.863 & 0.797 & 0.756 & 0.585 & 0.745 \\
			cautious2 & role-play   & 0.941 & 0.662 & 0.790 & 0.695 & 0.790 & 0.794 \\
			concise   & descriptive & 0.708 & 0.706 & 0.869 & 0.859 & 0.816 & 0.737 \\
			concise   & role-play   & 0.874 & 0.771 & 0.830 & 0.785 & 0.598 & 0.838 \\
			descriptive & role-play & 0.821 & 0.619 & 0.799 & 0.815 & 0.423 & 0.744 \\
			\bottomrule
		\end{tabular}
		\caption{Pairwise similarity scores across prompt variants for different language models.}
		\label{tab:pairwise_similarity}
	\end{table*}
	
	We also executed a similar ablation experiment as the one above, but instead of testing five different prompts, the models were given the original instruction prompt five times. The pairwise agreement scores for each run can be found in Table~\ref{tab:run_similarity}. It appears that our attempts to restrict stochasticity within the experiment (e.g., the standardization of python, numpy and pytorch seeds), as well as the nature of the task, lead to deterministic outputs by the LLMs on the evaluated sample.
	
	\begin{table}[t]
		\centering
		\small
		\begin{tabular}{llcccccc}
			\toprule
			\rotatebox[origin=c]{90}{\textbf{Run A}} & \rotatebox[origin=c]{90}{\textbf{Run B}} & \rotatebox[origin=c]{90}{\textbf{LLaMa-70B}} & \rotatebox[origin=c]{90}{\textbf{Qwen-32B}} & \rotatebox[origin=c]{90}{\textbf{OLMo-32B}} & \rotatebox[origin=c]{90}{\textbf{OLMo-7B}} & \rotatebox[origin=c]{90}{\textbf{Qwen-7B}} & \rotatebox[origin=c]{90}{\textbf{LLaMa-8B}} \\
			\midrule
			1 & 2 & 1 & 1 & 1 & 1 & 1 & 1 \\
			1 & 3 & 1 & 1 & 1 & 1 & 1 & 1 \\
			1 & 4 & 1 & 1 & 1 & 1 & 1 & 1 \\
			1 & 5 & 1 & 1 & 1 & 1 & 1 & 1 \\
			2 & 3 & 1 & 1 & 1 & 1 & 1 & 1 \\
			2 & 4 & 1 & 1 & 1 & 1 & 1 & 1 \\
			2 & 5 & 1 & 1 & 1 & 1 & 1 & 1 \\
			3 & 4 & 1 & 1 & 1 & 1 & 1 & 1 \\
			3 & 5 & 1 & 1 & 1 & 1 & 1 & 1 \\
			4 & 5 & 1 & 1 & 1 & 1 & 1 & 1 \\
			\bottomrule
		\end{tabular}
		\caption{Pairwise similarity scores across repeated runs for different language models using the original survey instructions.}
		\label{tab:run_similarity}
	\end{table}

	\subsection{One Model to Rule Them All?}
	\label{app:experiments:one-model}
	
	Since comments in the written and oral datasets are fundamentally different (\S\ref{app:dataset:exploratory}), we would be tempted to train different models for each. This decision is not obvious however, since it is likely that the model can internally distinguish between oral and written speech. If that occurs, the differences between the texts should not negatively affect the model; in fact, training the model on all datasets could allow it to transfer knowledge between the two text modalities. Table~\ref{tab:merged_metrics} shows the performance of the same model when trained only on oral and written datasets, compared to when trained on all datasets at once. We find that the models perform better when trained on different splits.
	
	\begin{table}[ht]
		\small
		\centering
		\begin{tabular}{lrrrr}
			\toprule
			Split & Dataset & Precision & Recall & F1 \\
			\midrule
			\multirow{2}{*}{Written} & CeRI      & 0.320 & 0.490 & 0.387 \\
			& Wikitactics & 0.407 & 0.717 & 0.520 \\
			\midrule
			\multirow{3}{*}{oral} & Fora      & 0.409 & 0.631 & 0.497 \\
			& IQ2       & 0.399 & 0.634 & 0.490 \\
			& WHoW      & 0.418 & 0.626 & 0.502 \\
			\midrule
			\multirow{4}{*}{All}   & CeRI      & 0.292 & 0.172 & 0.216 \\
			& Fora      & 0.410 & 0.664 & 0.507 \\
			& IQ2       & 0.406 & 0.623 & 0.491 \\
			& WHoW      & 0.422 & 0.642 & 0.509 \\
			& Wikitactics & 0.456 & 0.364 & 0.405 \\
			\bottomrule
		\end{tabular}
		\caption{Summary of classifier performance depending on the training dataset splits on the facilitator prediction task. We compare models trained on all available data with models trained separately on written and oral datasets. Performance is generally similar.}
		\label{tab:merged_metrics}
	\end{table}

	\subsection{Decision Thresholds}
	\label{app:experiments:thresholds}
	
	An advantage of using \ac{eo} classifiers instead of LLMs is that we can quantitatively alter the performance of the classifier by tuning the thresholds required for a comment to trigger an intervention. By making the model more conservative (increasing the threshold), we could hope that we can avoid unnecessary interventions, while making it less conservative (decreasing the threshold) we could make it more proactive.
	
\begin{figure*}[t]
		\includegraphics[width=\columnwidth]{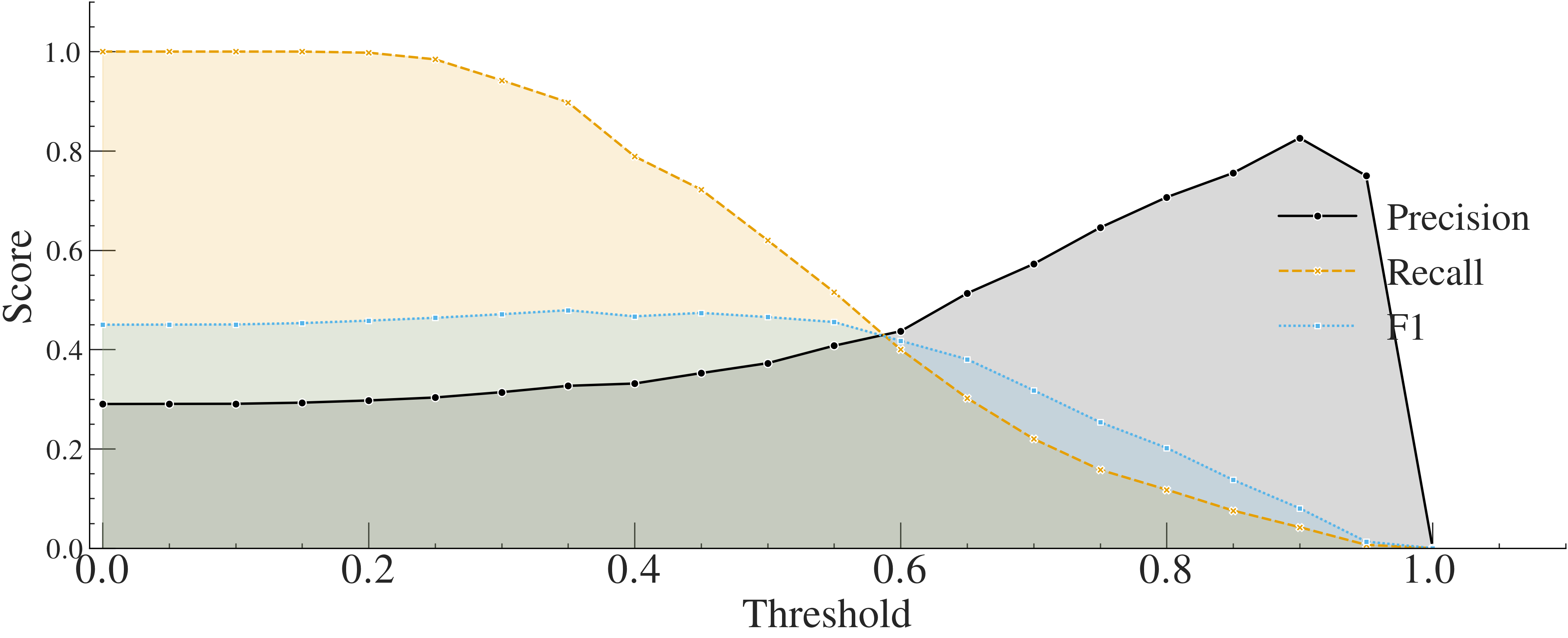}
		\includegraphics[width=\columnwidth]{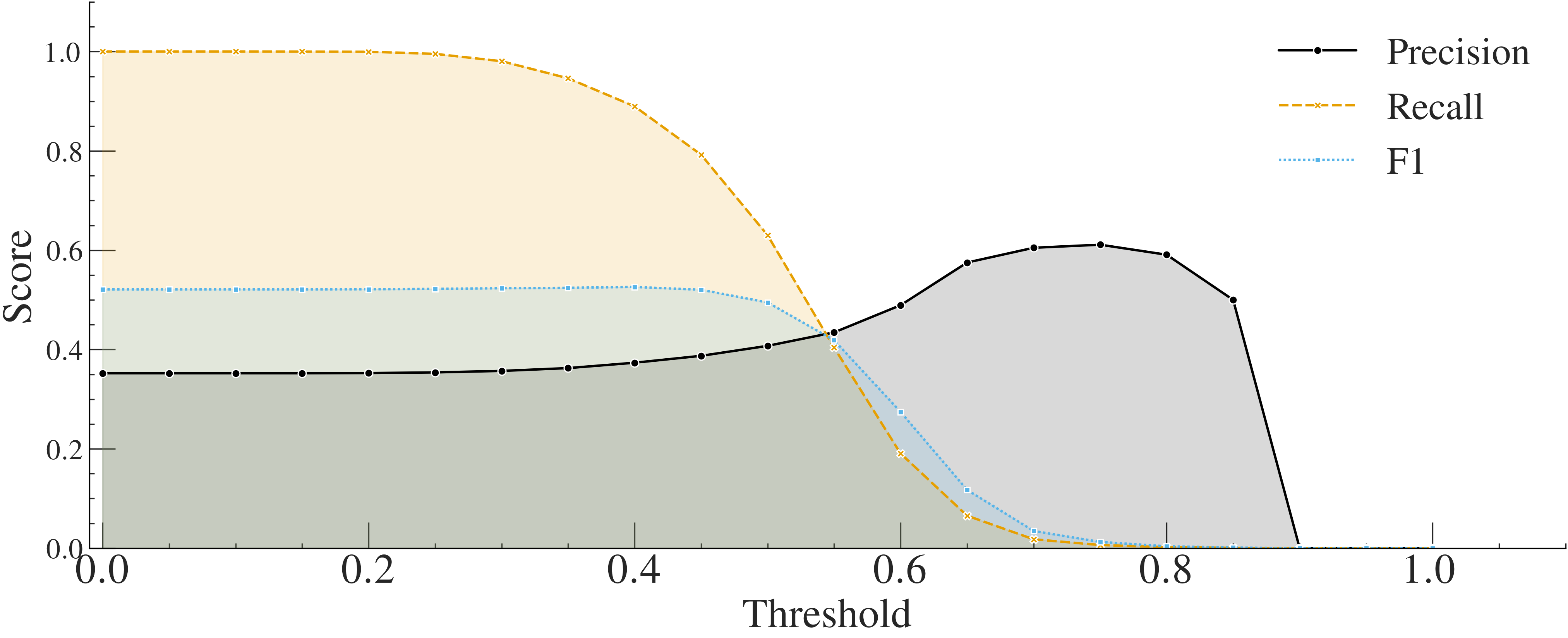}
		\caption{Classifier performance (\textbf{left}: written datasets, \textbf{right}: oral datasets) when tuning the decision threshold on the test-set. The higher the threshold, the more conservative the classifier is encouraged to become when predicting a facilitator intervention.}
		\label{fig:thresholds}
	\end{figure*}

	Fig.~\ref{fig:thresholds} shows that the classifier has an inherently low precision score which is largely invariant to the selected threshold. On the other hand, recall falls off rapidly, which incentivizes us to use a more liberal approach.
	
	\subsection{Full Results}
	
	Table~\ref{tab:facilitator-prediction-collapsed} shows the collapsed cases (F1$_p$ or F1$_n$ fall below 0.15)  of Table~\ref{tab:facilitator-prediction}.
	
	\begin{table}[ht]
		\centering
		\scriptsize
		\begin{tabular}{lrrrrr}
			\toprule
			Model & Pr. & Rec. & F1$_p$ & F1$_n$ & Sup.\\
			\midrule
			
			\multicolumn{6}{c}{\textbf{\ac{ceri}}} \\
			\midrule
			LLaMa70B & 0.152 & 0.082 & 0.107 & 0.783 & 335 \\
			OLMo32B  & 0.154 & 0.071 & 0.097 & 0.795 & 335 \\
			OLMo7B   & 0.251 & 0.988 & 0.401 & 0.000 & 335 \\
			Qwen32B  & 0.100 & 0.024 & 0.038 & 0.821 & 335 \\
			Qwen7B   & 0.182 & 0.118 & 0.143 & 0.774 & 335 \\
			{All} & {0.201} & {0.327} & {0.215} & {0.635} & {335} \\
			
			\midrule
			\multicolumn{6}{c}{\textbf{\ac{wt}}} \\
			\midrule
			LLaMa8B  & 0.328 & 0.842 & 0.472 & 0.074 & 336 \\
			OLMo7B   & 0.348 & 0.958 & 0.511 & 0.009 & 336 \\
			{All} & {0.307} & {0.581} & {0.393} & {0.361} & {336} \\
			
			\midrule
			\multicolumn{6}{c}{\textbf{Fora}} \\
			\midrule
			LLaMa70B & 0.333 & 0.058 & 0.099 & 0.760 & 382 \\
			OLMo32B  & 0.391 & 0.066 & 0.113 & 0.765 & 382 \\
			OLMo7B   & 0.358 & 0.993 & 0.526 & 0.008 & 382 \\
			Qwen32B  & 0.438 & 0.051 & 0.092 & 0.773 & 382 \\
			{All} & {0.390} & {0.355} & {0.279} & {0.604} & {382} \\
			
			\midrule
			\multicolumn{6}{c}{\textbf{\ac{iq2}}} \\
			\midrule
			OLMo32B  & 0.273 & 0.092 & 0.137 & 0.736 & 374 \\
			OLMo7B   & 0.345 & 0.977 & 0.510 & 0.000 & 374 \\
			Qwen32B  & 0.167 & 0.038 & 0.062 & 0.743 & 374 \\
			{All} & {0.305} & {0.409} & {0.306} & {0.534} & {374} \\
			
			\midrule
			\multicolumn{6}{c}{\textbf{\ac{whow}}} \\
			\midrule
			OLMo32B  & 0.455 & 0.079 & 0.135 & 0.779 & 364 \\
			OLMo7B   & 0.349 & 1.000 & 0.517 & 0.025 & 364 \\
			Qwen32B  & 0.296 & 0.063 & 0.105 & 0.762 & 364 \\
			{All} & {0.364} & {0.428} & {0.326} & {0.561} & {364} \\
			
			\bottomrule
		\end{tabular}
		\caption{Facilitator prediction results for collapsed models, where F1$_p$ or F1$_n$ fall below 0.15. Positive-class precision, recall and F1 (F1$_p$), and negative-class F1 (F1$_n$). Models that collapse on a dataset are listed only under that dataset.}
		\label{tab:facilitator-prediction-collapsed}
	\end{table}

	\section{Human Survey}
	\label{app:human-annotation}
	
	\subsection{Facilitative Training}
	\label{app:human-annotation:tactics}
	
	Before the main survey on facilitative intervention, all participants were extensively trained on currently established facilitative tactics. This task aimed to familiarize the participants with what constitutes a facilitative intervention, as well as how professional facilitators operate in real-life scenarios.
	
	Specifically, we split the participants into three groups, each of which was assigned a different taxonomy of facilitative interventions~\citep{chen-etal-2025-whow, schroeder-etal-2024-fora, ceri}. The participants were tasked with classifying 830 facilitated comments from a sample of continuous discussions in \ac{pefk}. The guidelines and results of this annotation are not shown in this study, but they are available upon request.
	
	We limited the training scope to strategies only, so that participants received no implicit or explicit input regarding the timing or flow of conversation. Training was introduced after preliminary discussions showed that participants tended to equate the task with content moderation. While training could in principle bias participants, restricting it to tactics was intended to keep the judgment of \emph{whether} and \emph{when} to intervene entirely their own.

	\subsection{Positive vs. Negative Reinforcement}
	\label{app:human-annotation:positive-vs-negative}
	
	\begin{table}[t]
		\centering
		\small
		\begin{tabular}{ll}
			\toprule
			\textbf{Taxonomy \& Tactic} & \textbf{Type} \\
			\midrule
			\multicolumn{2}{c}{\textbf{Fora}} \\
			\midrule
			Express appreciation & + \\
			Express agreement or affirmation & + \\
			Open invitation to participants &   \\
			Specific invitation to participant &   \\
			Model examples &   \\
			Follow-up question &   \\
			Make connections &   \\
			
			\midrule
			\multicolumn{2}{c}{\ac{whow}} \\
			\midrule
			Informational &   \\
			Coordinative & - \\
			Social & + \\
			Probing &   \\
			Confronting &   \\
			Instruction & - \\
			Interpretation &   \\
			Supplement &   \\
			
			\midrule
			\multicolumn{2}{c}{\ac{ceri}} \\
			\midrule
			Welcoming & + \\
			Encouragement / appreciation & + \\
			Maintaining civil discourse & - \\
			Explaining why comment not removed &   \\
			Indicating irrelevant/off-point comments & - \\
			Pointing to relevant information &   \\
			Pointing out effective commenting & + \\
			Asking for more information/sources &   \\
			Asking for solutions/alternatives &   \\
			Encourage engagement with others &   \\
			Pose question to community &   \\
			
			\bottomrule
		\end{tabular}
		\caption{Facilitation taxonomies (bold) with associated tactics (normal text) and reinforcement type. +: positive reinforcement, -: negative reinforcement, empty if the tactic does not clearly correspond to any category.}
	\end{table}
	
	The distinction between positive/negative reinforcement has been implicitly suggested in facilitative tactics taxonomies (Table~\ref{tab:datasets}), including prior work such as Fora~\cite{schroeder-etal-2024-fora} \ac{whow}~\citep{chen-etal-2025-whow}, and \ac{ceri}~\citep{ceri}. We conservatively estimate whether the tactics described in each taxonomy correspond to our definitions of positive and negative reinforcement. This approach also parallels the work of \citet{falk-etal-2021-predicting}, who use high and low Argument Quality for further analysis of facilitation prediction. Unlike their approach, we obtain the explanatory signal directly from the survey participants, rather than from established \ac{eo} models.
	
	The facilitative taxonomy proposed in Fora does not include any tactics that could be unambiguously considered as negative reinforcement, which is expected, since facilitators deployed in the relevant study were supposed to only act as coordinators. Notably, several tactics may simultaneously exhibit both positive and negative characteristics across all taxonomies, depending on context and the exact implementation used by the facilitator. We therefore use these taxonomies only as guidance during annotation, and refer the reader to the original works for a more comprehensive description and analysis of facilitation strategies.

	\subsection{Are Facilitative Tendencies Dataset-Dependent?}
	\label{app:human-annotation:datasets}
	
	\begin{figure}[t]
		\centering
		\includegraphics[width=\columnwidth]{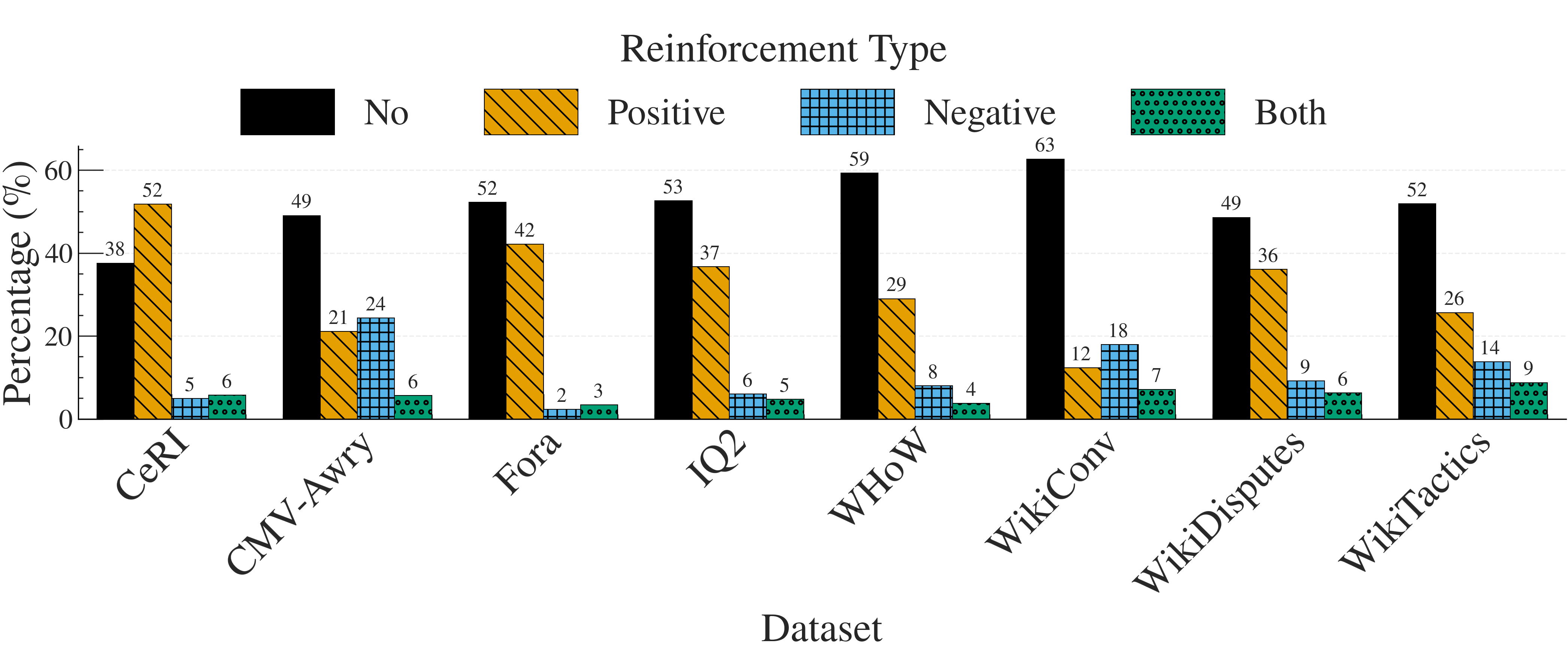}
		\caption{Label distributions of human participants per dataset.}
		\label{fig:label_distribution_dataset}
	\end{figure}
	
	Facilitative tendencies generally follow a conservative pattern, with no reinforcement selected most often. Despite the differences between the datasets (forums, debates, and radio podcasts), we observe similar trends in the choice between positive and negative reinforcement, with positive reinforcement being more common overall. Two exceptions are cmv\_awry and wikiconv, where negative interventions are selected more often than positive ones, by 3\% and 6\%, respectively, and ceri, where positive reinforcement is selected more often than no reinforcement (Fig.~\ref{fig:label_distribution_dataset}).

	\subsection{Participant Consistency}
	\label{app:human-annotation:intra-agreement}
	
	To evaluate the consistency of individual participants, we measured intra-participant agreement using purposefully repeated instances of identical discussions. 
	
	Agreement was computed as a strict match in the labeling of the same instance (duplicated pair). For each duplicated pair, agreement was assigned a value of 1 if both labels were identical and 0 otherwise in order to calculate a simple pairwise agreement. Hence, the final intra-participant agreement for each participant was then calculated as the mean of these binary agreement scores. Since our classes are heavily skewed and imbalanced, traditional agreement metrics tend to collapse, showing extremely low scores even for almost-100\% agreement~\citep{rottger-etal-2022-two, Feinstein1990HighAgreement}. For the purposes of simplicity, we thus run our analysis in this part with simple percentage agreement.
	
	We report results in Table~\ref{tab:intra_agreement}. Most participants demonstrated near-perfect consistency, with several achieving 100\% agreement. Participant A3 is the only participant who showed reduced consistency (66.67\%), suggesting potential variability in interpretation or labeling behavior.
	
	\begin{table}[t]
		\centering
		\begin{tabular}{cc}
			\toprule
			\textbf{participant} & \textbf{Agreement (\%)} \\
			\toprule
			A2  & 100.00 \\
			A3  & 66.67 \\
			A4  & 88.89 \\
			A5  & 100.00 \\
			A6  & 100.00 \\
			A7  & 100.00 \\
			A8  & 100.00 \\
			A9  & 100.00 \\
			A10 & 100.00 \\
			A11 & 100.00 \\
			\bottomrule
		\end{tabular}
		\caption{Intra-participant percentage agreement (consistency) over all subdatasets of \ac{pefk}.}
		\label{tab:intra_agreement}
	\end{table}

	\subsection{Participants Selection and Coordination}
	\label{app:participants_select}
	
	\paragraph{Selection} Participant selection required a response to the public call, followed by a round of interviews to verify that candidates fulfilled the criteria for the role. All successful participants held a BA in Languages, Communication, or History, and an MSc in either Digital Humanities or Language Technology. They also had experience with NLP annotation tasks, such as sentiment analysis and toxicity detection. We would also like to highlight that controls were conducted before the survey (hiring), during the survey (training and discussion), and after the survey (removal of outlier or inconsistent participants), to ensure the highest quality of results. The gender distribution was relatively balanced, with 4 men and 7 women, and the median age was 30 years. Before the commencement of the survey, participants completed an informed consent form in compliance with GDPR~\citep{GDPR} ethical requirements (App.~\ref{app:consent}) and were compensated in accordance with national legislation.
	
	\paragraph{Coordination} The coordination of the project was largely based on weekly meetings that allowed the integration of participant feedback and potential changes to our survey methodology. An overview of the project timeline can be found in Table~\ref{tab:timeline}. Before the commencement of the main survey, two important preparatory steps were carried out. The first was familiarizing participants with facilitation tactics; more information can be found in Section~\ref{app:human-annotation:tactics}. 
	
\begin{figure*}
		\centering
		\includegraphics[width=1\linewidth]{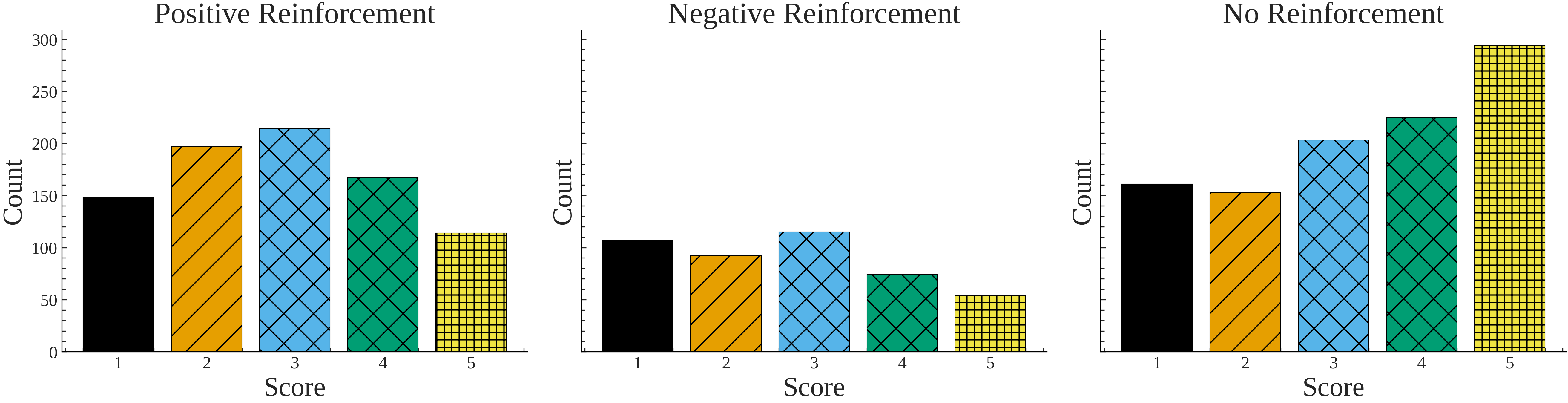}
		\caption{Frequencies of each reinforcement class on a Likert scale for the two pilot studies conducted before the main survey task.}
		\label{fig:instances_pilot}
	\end{figure*}

	The second was a pilot round with $\sim$160 instances, which allowed us to better understand the challenges of the task. For example, one major concern was time. Depending on instance length, labeling took between 2 and 5 minutes. The pilot helped us pin down to how much time the annotators needed. This led us to break the task into smaller subtasks, spread across almost 3 months (see Phases 5 and 6 in Table~\ref{tab:timeline}), with annotators labeling about 400 instances per round. Moreover, during this pilot, the annotators noticed coding errors, as well as malformed instances, which we addressed by adding malformed-instance flagging as an option. Finally, the pilot helped us refine the definitions of positive and negative reinforcement. For example, how to handle `yes' as a comment, which we considered positive if it had a positive impact on the conversation and was not used merely as hedging. Figure~\ref{fig:instances_pilot} shows the frequencies of each class on a Likert scale for the pilot. Because of the malformed instances in the initial dataset, and because the task itself changed between the pilot and the final survey, the two are not directly comparable. We nevertheless release the raw pilot survey files in the Supporting Material.
	
	\begin{table}[!t]
		\centering
		\footnotesize
		\setlength{\tabcolsep}{3pt}
		\begin{tabularx}{\columnwidth}{@{} l l X @{}}
			\toprule
			\textbf{Ph.} & \textbf{Dates} & \textbf{Activity} \\
			\midrule
			1 & Dec & Recruitment \\
			2 & Dec--Jan & Training on facilitation strategies \\
			3 & Jan 1--9 & Onboarding \\
			4 & Jan 9--15 & Pilot \\
			5 & Mid Jan--Mid Feb & Main experiment and discussion \\
			6 & End Feb--Mid Mar & Main experiment completion \\
			1--6 & Weekly & Discussion \& feedback rounds \\
			\bottomrule
		\end{tabularx}
		\caption{Coordination timeline.}
		\label{tab:timeline}
	\end{table}

	\subsection{Full Results of Human vs. LLM Disagreement}
	\label{app:sec:disagreement_examples}
	
	\begin{table}[t]
		\small
		\centering
		\begin{tabular}{lrrrr}
			\toprule
			& \multicolumn{2}{c}{\textbf{LLM annotators}} & \multicolumn{2}{c}{\textbf{Human annotators}} \\
			\cmidrule(lr){2-3} \cmidrule(lr){4-5}
			\textbf{Label} & $n$ & \% & $n$ & \% \\
			\midrule
			Positive & 113 & 52.3 &  19 &  5.4 \\
			Negative &  49 & 22.7 & 104 & 29.6 \\
			Both     &  42 & 19.4 &  13 &  3.7 \\
			No       &  12 &  5.6 & 215 & 61.3 \\
			\midrule
			Total    & 216 & 100.0 & 351 & 100.0 \\
			\bottomrule
		\end{tabular}
		\caption{Distribution of labels assigned by LLM and human annotators across the 36 extreme-disagreement cases. LLM labels come from six models. 9 missing (NaN) labels are excluded, leaving 351 labels in total. LLMs predominantly assign \textit{Positive}, whereas humans predominantly assign \textit{No}.}
		\label{tab:label_distribution_36}
	\end{table}
	
	Across all these 36 extreme disagreement cases, the LLM majority label never matches the human majority label. The percentages of the labels per class, for humans and LLMs individually can be found in Table~\ref{tab:label_distribution_36}. LLMs mostly say Positive (52\%), while humans mostly say No (61\%), followed by Negative (30\%). In many cases, the disagreement emerges due to the fact that LLMs `over-read' meaning, even in procedural or administrative chunks of discourse. These cases, in reality, have almost no content, like greetings and affirmations (``Thank you'', ``Yes'', ``Hi''), or a moderator managing turns (``All right. We get it''). Humans mostly labeled these cases with No, while the LLMs gave Positive or Both instead, as if every dialogue needed a substantive label. 
	
	Some other cases involve a person sustaining a lengthy argument with no real interaction with other interlocutors. Some topic examples include an Aviation High School story, the organ-market, and cyber-attacks. Despite taking the form of a narrative, LLMs label them as cases where Positive Reinforcement could help. Humans maintained that no intervention was needed. 
	
	The most concerning issue is a recall problem for possibly Negative reinforcement instances. Humans label threads with condescension or hostility as Negative, for example ``You are simply wrong'' and ``p\censor{is}s poor argument''. The latter example is about a debate about gender roles and childcare, where one user replies with a crude, graphic description of pregnancy aimed at the other person, who answers by calling it a ``p\censor{is}s poor argument''. All ten human annotators, plus \texttt{LLaMa8b} and \texttt{LLaMa70b}, argued for Negative Reinforcement. \texttt{Qwen32b}, \texttt{OLMo7b} and \texttt{OLMo32b} labeled it Positive, and \texttt{Qwen7b} labeled it Both. It is likely that models respond to the fact that arguments and counterpoints are being exchanged and overlook the hostility in how they're delivered.

	\subsection{Malformed Instances}
	
	\begin{figure}[t]
		\centering
		\includegraphics[width=\linewidth]{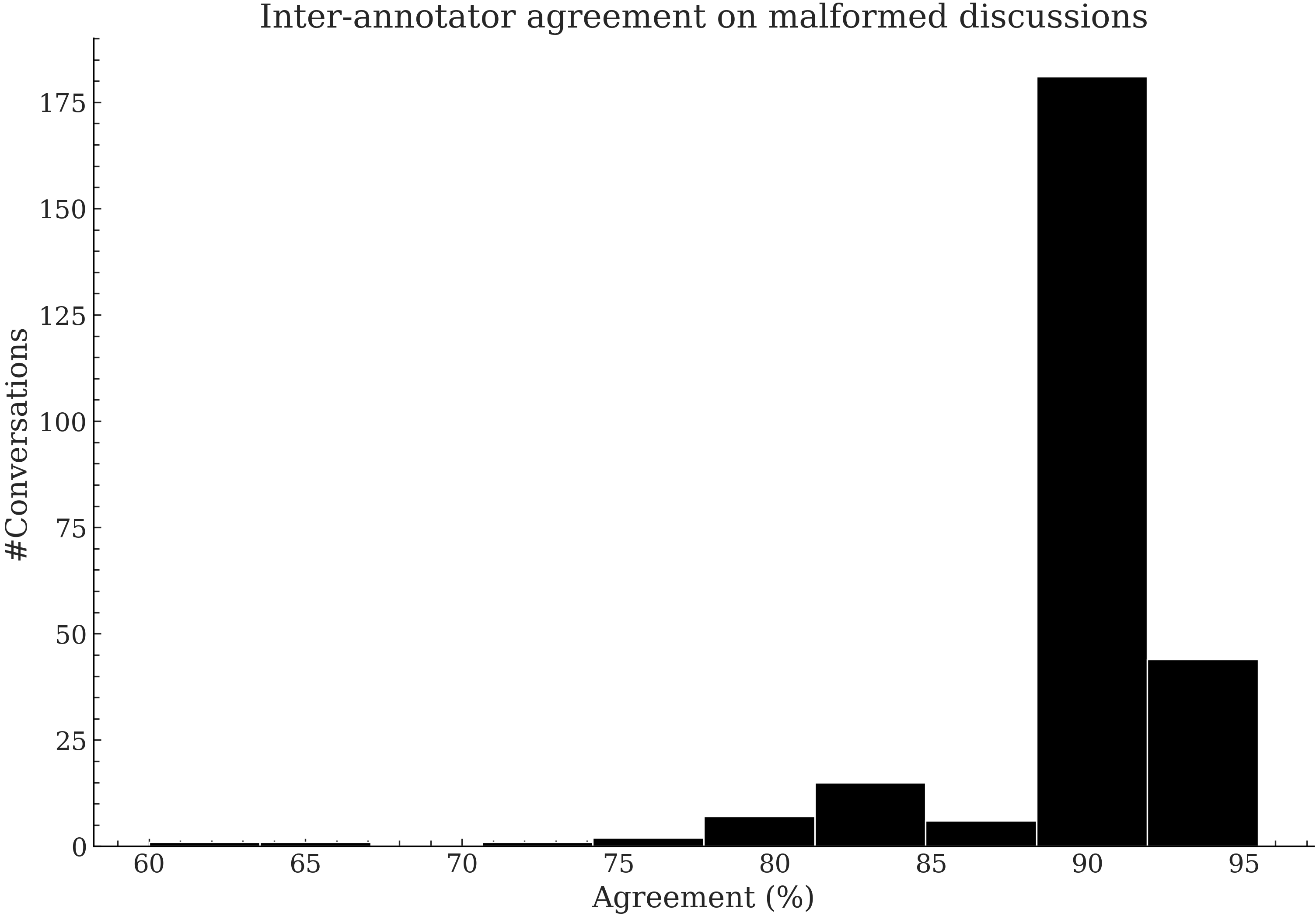}
		\caption{Agreement on malformed instances.}
		\label{fig:malformed}
	\end{figure}
	
	The participants were also required to report any malformed instances that might obscure the processing of the conversation (see \S\ref{app:guidelines}). The instances most consistently reported fall into two broad types.
	The first type is broken transcripts of spoken debates. These are chains of interrupted turns where no speaker finishes a sentence (e.g., ``It's an incredibly rare event---'' / ``But anyhow---'' / ``There was a---''; or ``We can get each of you --'' / ``I think it's important because they keep raising this issue of --'' / ``Well --''). Turns are often split mid-sentence across speakers (e.g., ``So I-- the common sense side of that argument [\ldots] Either of you-- either of you would take that.''), and they can contain empty placeholder turns rendered as \texttt{nan} or speech marked \texttt{[inaudible]} so often that the exchange can't be followed (e.g., ``You know [inaudible] we got to do before [inaudible] at the time''). Annotators described these as ``Sentences are truncated'' or simply ``Makes no sense''.
	The second type is online public-comment threads, such as consultations on transit funding, consumer debt collection, or federal health IT policy. Here standalone comments are split into ``turns'' and interleaved with templated moderator replies (e.g., ``Welcome and thank you for sharing your story''). They come with leftover HTML entities (e.g., \texttt{\&\#39;}, \texttt{\&nbsp;}, \texttt{\&quot;}) and encoding errors (e.g., \texttt{â€œ}, \texttt{â€™}, \texttt{\_x009d\_}) from character-set conversion. In both types there is little or no interaction between participants to judge.

	\subsection{Survey Guidelines} 
	\label{app:guidelines}
	
	\textbf{Survey Guidelines: Type of Reinforcement Needed in Comment Threads}
	
	\textbf{Purpose of the Task}
	
	These guidelines explain how to rate comments in conversation threads using three parallel rating columns:
	
	\begin{itemize}
		\item Positive Reinforcement (1--5)
		\item Negative Reinforcement (1--5)
		\item No Reinforcement Needed (1--5)
	\end{itemize}
	
	Your job is to read each comment in its conversational context and assign a strength/certainty rating (1--5) for each category, reflecting how that category applies. All fields need to be filled and the categories are not necessarily mutually exclusive.
	
	Even though your intervention will be delivered at the end of the thread, the reinforcement can refer to any point in the conversation, not just at the end. It can refer to something said earlier, not just the last comment.
	
	You are not being asked whether that conversation correlates to positive or negative reinforcement, but rather whether you, personally, would have used positive or negative reinforcement at a given point. Keep in mind that too many interventions on the moderator’s part may result in an adverse reaction from the participants; no one wants to be constantly told what to say or how to say it.
	
	\textbf{Important Notes}
	\begin{itemize}
		\item You must add the rationale behind your response in the rational/comments field when:
		\begin{itemize}
			\item your positive score $>$ 3, or
			\item your negative score $>$ 3, or
			\item you encounter a particular case (e.g., an instance which could fit both negative and positive reinforcement categories).
		\end{itemize}
		\item If you encounter data that is malformed or noisy to the point that it renders comprehension difficult, flag the instance using the malformed field.
	\end{itemize}
	
	\textbf{What Each Label Means}
	
	\textbf{Positive Reinforcement}
	
	Positive reinforcement is defined as an intervention whose purpose is to encourage a specific behavior. For example, a moderator may ask people to participate or ask a person to elaborate. Positive reinforcement is not related to positive speech; the moderator may be strict or even rude when attempting to encourage specific behaviors (e.g., ``I will not tolerate any further discussion on this topic. It's been decided and we're moving on.'').
	
	Rate this when a conversation demonstrates behavior that should be encouraged or reinforced in a synchronous exchange. Your task is not simply to identify positive behaviors, but to identify moments where reinforcing them would help sustain or build constructive conversational momentum.
	
	This includes, but is not limited to:
	\begin{itemize}
		\item Helpful or clarifying explanations
		\item Respectful disagreement
		\item Constructive feedback
		\item Polite, empathetic, or supportive tone
		\item Efforts to de-escalate conflict
		\item Any contribution that meaningfully improves the quality or direction of the conversation
	\end{itemize}
	
	\textbf{Negative Reinforcement}
	
	The purpose of negative reinforcement is to discourage unwanted or disruptive behaviors. This is correlated, but not necessarily closely related to threats or enforcement of disciplinary action. Negative reinforcement is not related to negative speech; the moderator may be kind and diplomatic when discouraging behaviors (e.g., ``I've received multiple complaints about your behavior on this forum. Please refrain from making personal attacks and inflammatory comments.'').
	
	Rate this when the comment reflects harmful, disruptive, or inappropriate behavior that should be discouraged, as well as cases where some participants dominate the conversation, or when the conversation starts to go off topic. Other examples include:
	\begin{itemize}
		\item Hostile, dismissive, or disrespectful language
		\item Personal attacks or insults
		\item Repeated interruptions or monopolizing the exchange
		\item Escalating conflict rather than resolving it
		\item Contributions that reduce the overall quality of the conversation
	\end{itemize}
	
	\textbf{No Reinforcement Needed}
	
	Rate this when the comment is neutral, unremarkable, or does not affect the tone or direction of the conversation. This includes:
	\begin{itemize}
		\item Basic factual statements
		\item Clarifying questions
		\item Minimal or vague comments
		\item Harmless but non-notable contributions
	\end{itemize}
	
	\textbf{How to Make Decisions}
	
	\textbf{Look at Intent and Impact}
	
	Consider what the commenter seems to be trying to do, and how their words might affect others in the thread.
	
	\textbf{Use the Context}
	
	Always check the surrounding messages. A comment that looks fine on its own might escalate tension when viewed in context—and vice versa.
	
	\textbf{Judge Behavior, Not Opinions}
	
	The label should never depend on whether you personally agree with the comment. Focus on tone, clarity, constructiveness, and how it influences the conversation.
	
	\textbf{Remain Consistent}
	
	Apply the same standards across all comments. Avoid guessing or over-interpreting; use what is present in the text.

	\subsection{Consent Forms}
	\label{app:consent}
	
	We include an abridged version of the informed consent form signed by the survey participants. For the full text, contact the \textit{``Athena'' Research Center Ethics Committee (EHDE)} mentioning Protocol Authorization Number \texttt{66-17/11/2025}.
	
	\begin{description}
		\item[Eligibility and Tasks:] Participants were selected based on their educational background and familiarity with Humanities, Artificial Intelligence (AI), and Natural Language Processing (NLP). Tasks included multiple-choice questionnaires, open-ended questions, and short interviews.
		
		\item[Benefits and Risks:] Participants were informed of professional benefits, including practical experience in data labeling and contributing to the development of tools to combat online hate speech, racism, and toxicity. Potential risks included a significant time commitment and exposure to toxic language.
		
		\item[Data Privacy and Confidentiality:] The study ensured anonymity by assigning participants random IDs and utilizing encrypted, non-identifiable data storage. No personal information (e.g., names, IP addresses) is stored online. Data may be used for academic publications and machine learning model training in a summarized, anonymized form.
		
		\item[Ethics and Compensation:] The research was funded by the ``Archimedes'' project and received approval from the Ethics and Deontology Committee of the Athena Research Center. Compensation is determined by the participants' individual contracts with the ``Archimedes'' unit. 
		
		\item[Reporting:] Participants were provided with contact information for the research team, the Ethics Committee, and the Data Protection Officer (DPO) for any concerns or complaints regarding the study or data handling.
	\end{description}

	\section{LLM Prompts}
	\label{app:guidelines:llm}
	
	\subsection{Main Prompts}
	
	\subsubsection{Yes/No Intervention}
	\label{app:guidelines:llm:yes-no}
	
	You are an expert specializing in online moderation and facilitation.
	You will be presented with a comment and the comments made before it.
	Choose whether you decide to intervene at this point in the discussion.
	
	Key Rules: 
	
	\textbf{Look at Intent and Impact}
	Consider what the commenter seems to be trying to do, and how their words might affect others in the thread.
	
	\textbf{Use the Context}
	Always check the surrounding messages. A comment that looks fine on its own might escalate tension when viewed in context—and vice versa. 
	
	\textbf{Judge Behavior, Not Opinions}
	The label should never depend on whether you personally agree with the comment. Focus on tone, clarity, constructiveness, and how it influences the conversation.
	
	\textbf{Remain Consistent}
	Apply the same standards across all comments. Avoid guessing or over-interpreting; use what is present in the text.
	
	\textbf{Output}
	Respond whether you would intervene in this discussion as a facilitator on a scale of 1-5, where 1 is "I'm sure I don't need to intervene", and 5 is "I'm sure I need to intervene"

	\subsubsection{Positive vs. Negative reinforcement}
	\label{app:guidelines:llm:positive-vs-negative}
	
	You are an expert facilitator. You will be presented with a discussion. Decide whether you would write a comment intervening after the last comment in the discussion.
	
	\textbf{The Core Rating Task}
	
	For every comment, you must write out a response of 1 (uncertain) to 5 (certain):

	\textbf{Positive Reinforcement}
	
	Defined as an intervention whose purpose is to encourage a specific behavior. For example, a moderator may ask people to participate, ask a person to elaborate etc.
	
	Positive reinforcement is not related to positive speech; the moderator may be strict or even rude when attempting to encourage specific behaviors (e.g., ``I will not tolerate any further discussion on this topic. It's been decided and we're moving on.'').
	
	Rate this when a conversation demonstrates behavior that should be encouraged or reinforced in a synchronous exchange. Your task is not simply to identify positive behaviors, but to identify moments where reinforcing them would help sustain or build constructive conversational momentum.
	
	This includes, but is not limited to:
	\begin{itemize}
		\item Helpful or clarifying explanations
		\item Respectful disagreement
		\item Constructive feedback
		\item Polite, empathetic, or supportive tone
		\item Efforts to de-escalate conflict
		\item Any contribution that meaningfully improves the quality or direction of the conversation
	\end{itemize}
	
	\textbf{Negative Reinforcement}
	
	The purpose of negative reinforcement is to discourage unwanted or disruptive behaviors. This is correlated, but not necessarily closely related to threats or enforcement of disciplinary action.
	
	Negative reinforcement is not related to negative speech; the moderator may be kind and diplomatic when discouraging behaviors (e.g., ``I've received multiple complaints about your behavior on this forum. Please refrain from making personal attacks and inflammatory comments.'').
	
	Rate this when the comment reflects harmful, disruptive, or inappropriate behavior that should be discouraged, as well as cases where some participants dominate the conversation, or when the conversation starts to go off topic.
	
	Other examples include:
	\begin{itemize}
		\item Hostile, dismissive, or disrespectful language
		\item Personal attacks or insults
		\item Repeated interruptions or monopolizing the exchange
		\item Escalating conflict rather than resolving it
		\item Contributions that reduce the overall quality of the conversation
	\end{itemize}
	
	\textbf{No Reinforcement}
	
	Rate this when the comment is neutral, unremarkable, or does not affect the tone or direction of the conversation.
	
	This includes:
	\begin{itemize}
		\item Basic factual statements
		\item Clarifying questions
		\item Minimal or vague comments
		\item Harmless but non-notable contributions
	\end{itemize}
	
	\textbf{Key Rules:}
	
	\textbf{Look at Intent and Impact} \\
	Consider what the commenter seems to be trying to do, and how their words might affect others in the thread.
	
	\textbf{Use the Context} \\
	Always check the surrounding messages. A comment that looks fine on its own might escalate tension when viewed in context—and vice versa.
	
	\textbf{Judge Behavior, Not Opinions} \\
	The label should never depend on whether you personally agree with the comment. Focus on tone, clarity, constructiveness, and how it influences the conversation.
	
	\textbf{Remain Consistent} \\
	Apply the same standards across all comments. Avoid guessing or over-interpreting; use what is present in the text.
	
	\textbf{Output}
	
	Respond whether you would intervene in this discussion as a facilitator.
	
	Always respond with the following format and nothing else:
	
	\begin{verbatim}
		Positive: 1-5
		Negative: 1-5
		No Reinforcement: 1-5
	\end{verbatim}

	\subsubsection{Single-Comment Facilitation Prediction}
	\label{app:guidelines:llm:timing-prediction}
	
	You are an expert moderator/facilitator.
	You will be presented with a comment (TGT) and the comments made before it (CTX, one for each comment in turn).
	Choose whether you decide to intervene at this point in the discussion.

	Key Rules:
	
	\textbf{Look at Intent and Impact}
	Consider what the commenter seems to be trying to do, and how their words might affect others in the thread.
	
	\textbf{Use the Context}
	Always check the surrounding messages. A comment that looks fine on its own might escalate tension when viewed in context—and vice versa.
	
	\textbf{Judge Behavior, Not Opinions}
	The label should never depend on whether you personally agree with the comment. Focus on tone, clarity, constructiveness, and how it influences the conversation.
	
	\textbf{Remain Consistent}
	Apply the same standards across all comments. Avoid guessing or over-interpreting; use what is present in the text.
	
	\textbf{Output}
	Respond whether you would intervene in this discussion as a facilitator. Write 0 if you would not intrevene, and 1 if you would. Do not write anything else, only the number.

	\subsubsection{Single-Comment Facilitation Detection}
	\label{app:guidelines:llm:timing-detection}
	
	You are an expert moderator/facilitator.
	You will be presented with a comment (TGT) and the comments made before it (CTX, one for each comment in turn).
	Choose whether the comment you are reading (TGT) is a comment made by a moderator/facilitator.

	Key Rules for:
	
	\textbf{Look at Intent and Impact}
	Consider what the commenter seems to be trying to do, and how their words might affect others in the thread.
	
	\textbf{Use the Context}
	Always check the surrounding messages. A comment that looks fine on its own might escalate tension when viewed in context—and vice versa.
	
	\textbf{Judge Behavior, Not Opinions}
	The label should never depend on whether you personally agree with the comment. Focus on tone, clarity, constructiveness, and how it influences the conversation.
	
	\textbf{Remain Consistent}
	Apply the same standards across all comments. Avoid guessing or over-interpreting; use what is present in the text.
	
	\textbf{Output}
	Respond whether the comment you are reading (TGT) is a comment made by a moderator/facilitator. Write 0 if it was not made by a facilitator, and 1 if it was made by a facilitator. Do not write anything else, only the number.

	\subsection{Ablation Prompts}
	\label{app:guidelines:llm:ablation}
	
	\subsubsection{Cautious1}
	You are an expert facilitator.
	You will be presented with a discussion.
	Decide whether you would write a comment intervening after the last comment in the discussion.
	
	\subparagraph{The Core Rating Task}
	
	For every discussion, produce three ratings from 1 (uncertain) to 5 (certain).
	
	\textbf{Positive Reinforcement}
	
	Defined as an intervention whose purpose is to encourage a specific behavior. For example, a facilitator may ask people to participate, ask a person to elaborate, or encourage constructive discussion. Positive reinforcement is not related to positive sentiment; the facilitator may be strict or direct when encouraging specific behaviors (e.g., ``I will not tolerate any further discussion on this topic. It's been decided and we're moving on.'').
	
	Assign this rating when the conversation demonstrates behavior that should be encouraged or reinforced in a synchronous exchange. Your task is not simply to identify positive behaviors, but to identify moments where reinforcing them would help sustain or build constructive conversational momentum.
	
	Examples include, but are not limited to:
	\begin{itemize}
		\item Helpful or clarifying explanations
		\item Respectful disagreement
		\item Constructive feedback
		\item Polite, empathetic, or supportive tone
		\item Efforts to de-escalate conflict
		\item Contributions that meaningfully improve the quality or direction of the conversation
	\end{itemize}
	
	\textbf{Negative Reinforcement}
	
	The purpose of negative reinforcement is to discourage unwanted or disruptive behaviors. This is correlated with, but not limited to, disciplinary action. Negative reinforcement is not related to negative sentiment; the facilitator may be kind and diplomatic when discouraging behaviors (e.g., ``I've received multiple complaints about your behavior on this forum. Please refrain from making personal attacks and inflammatory comments.'').
	
	Assign this rating when the conversation reflects harmful, disruptive, or inappropriate behavior that should be discouraged, as well as cases where participants dominate the conversation or the discussion starts to drift off topic.
	
	Examples include:
	\begin{itemize}
		\item Hostile, dismissive, or disrespectful language
		\item Personal attacks or insults
		\item Repeated interruptions or monopolizing the exchange
		\item Escalating conflict rather than resolving it
		\item Contributions that reduce the overall quality of the conversation
	\end{itemize}
	
	\textbf{No Reinforcement}
	
	Assign this rating when the conversation is neutral, unremarkable, or does not meaningfully affect the tone or direction of the discussion. This includes:
	\begin{itemize}
		\item Basic factual statements
		\item Clarifying questions
		\item Minimal or vague comments
		\item Harmless but otherwise unremarkable contributions
	\end{itemize}
	
	Remember that the role of the facilitator is supportive rather than dominant. While intervention is sometimes necessary, unnecessary or frequent intervention degrades the user experience.
	
	\subparagraph{Key Rules}
	
	\textbf{Look at Intent and Impact}
	
	Consider what the commenter seems to be trying to do, and how their words might affect others in the thread.
	
	\textbf{Use the Context}
	
	Always consider the surrounding discussion. A comment that appears appropriate in isolation may escalate tension in context, and vice versa.
	
	\textbf{Judge Behavior, Not Opinions}
	
	The labels should never depend on whether you personally agree with the comment. Focus on tone, clarity, constructiveness, and how the discussion evolves.
	
	\textbf{Remain Consistent}
	
	Apply the same standards across all discussions. Avoid guessing or over-interpreting; evaluate only what is present in the conversation.
	
	\subparagraph{Output}
	
	Respond only using the following format and nothing else:
	
	\begin{verbatim}
		Positive: 1-5
		Negative: 1-5
		No Reinforcement: 1-5
	\end{verbatim}

	\subsubsection{Cautious2}	
	You are an expert facilitator.
	You will be presented with a discussion.
	Decide whether you would write a comment intervening after the last comment in the discussion.
	
	\subparagraph{The Core Rating Task}
	
	For every discussion, provide three ratings on a scale from 1 (uncertain) to 5 (certain).
	
	\textbf{Positive Reinforcement}
	
	Positive reinforcement refers to an intervention whose purpose is to encourage a specific behavior. For example, a facilitator may encourage participation, ask someone to elaborate, or reinforce constructive conversational habits. Positive reinforcement is not synonymous with positive sentiment; the facilitator may be strict or direct when encouraging desired behaviors (e.g., ``I will not tolerate any further discussion on this topic. It's been decided and we're moving on.'').
	
	Assign this rating when the discussion demonstrates behavior that should be encouraged or reinforced in a synchronous exchange. Do not simply identify positive behaviors; instead, identify situations where reinforcing them would help sustain or strengthen constructive conversational momentum.
	
	Examples include, but are not limited to:
	\begin{itemize}
		\item Helpful or clarifying explanations
		\item Respectful disagreement
		\item Constructive feedback
		\item Polite, empathetic, or supportive communication
		\item Attempts to de-escalate conflict
		\item Contributions that meaningfully improve the quality or direction of the discussion
	\end{itemize}
	
	\textbf{Negative Reinforcement}
	
	Negative reinforcement refers to an intervention whose purpose is to discourage unwanted or disruptive behaviors. It may involve setting boundaries or discouraging problematic conduct, but does not necessarily involve threats or disciplinary action. Likewise, it is not synonymous with negative sentiment; a facilitator may politely discourage undesirable behavior (e.g., ``I've received multiple complaints about your behavior on this forum. Please refrain from making personal attacks and inflammatory comments.'').
	
	Assign this rating when the discussion contains harmful, disruptive, or inappropriate behavior that should be discouraged, including situations where participants dominate the conversation or the discussion begins to drift off topic.
	
	Examples include:
	\begin{itemize}
		\item Hostile, dismissive, or disrespectful language
		\item Personal attacks or insults
		\item Repeated interruptions or monopolizing the discussion
		\item Escalating conflict rather than resolving it
		\item Contributions that reduce the overall quality of the conversation
	\end{itemize}
	
	\textbf{No Reinforcement}
	
	Assign this rating when the discussion is neutral, unremarkable, or does not meaningfully affect the tone or direction of the conversation. This includes:
	\begin{itemize}
		\item Basic factual statements
		\item Clarifying questions
		\item Minimal or vague comments
		\item Harmless but otherwise unremarkable contributions
	\end{itemize}
	
	Your intervention ratings should prioritize \textbf{surgical precision}. Over-moderation is counterproductive, and unnecessary interventions will degrade the participant experience.
	
	\textbf{Key Rules}
	
	\textbf{Look at Intent and Impact}
	
	Consider what each commenter appears to be trying to accomplish, and how their contribution is likely to affect the discussion.
	
	\textbf{Use the Context}
	
	Always evaluate comments in the context of the surrounding discussion. A comment that appears appropriate in isolation may escalate tension in context, and vice versa.
	
	\textbf{Judge Behavior, Not Opinions}
	
	Your ratings should never depend on whether you personally agree with the content of a comment. Focus instead on tone, clarity, constructiveness, and the comment's effect on the conversation.
	
	\textbf{Remain Consistent}
	
	Apply the same standards across all discussions. Avoid over-interpreting intent or inferring information that is not present.
	
	\subparagraph{Output}
	
	Respond \emph{only} with the following format:
	
	\begin{verbatim}
		Positive: 1-5
		Negative: 1-5
		No Reinforcement: 1-5
	\end{verbatim}
	
	\subsubsection{Concise}
	You are an expert facilitator.
	You will be presented with a discussion.
	Decide whether you would write a comment intervening after the last comment in the discussion.
	
	\subparagraph{The Core Rating Task}
	
	For every discussion, provide three ratings on a scale from 1 (low intensity) to 5 (high intensity).
	
	\textbf{Positive Reinforcement}
	
	Rate highly if the discussion demonstrates constructive behaviors that should be encouraged or reinforced. Examples include:
	\begin{itemize}
		\item Helpful explanations or elaboration
		\item Respectful disagreement
		\item Constructive feedback
		\item Polite, empathetic, or supportive communication
		\item Attempts to de-escalate conflict
		\item Contributions that improve the quality or direction of the discussion
	\end{itemize}
	
	This rating reflects opportunities where facilitator reinforcement would strengthen productive conversational dynamics.
	
	\textbf{Negative Reinforcement}
	
	Rate highly if the discussion contains behaviors that should be discouraged. Examples include:
	\begin{itemize}
		\item Hostile, dismissive, or disrespectful language
		\item Personal attacks or insults
		\item Repeated interruptions or monopolizing the discussion
		\item Escalating conflict rather than resolving it
		\item Going substantially off topic
		\item Contributions that reduce the overall quality of the discussion
	\end{itemize}
	
	This rating reflects opportunities where facilitator intervention would discourage disruptive conversational dynamics.
	
	\textbf{No Reinforcement}
	
	Assign a high rating when the discussion is largely neutral and does not warrant intervention. Examples include:
	\begin{itemize}
		\item Basic factual statements
		\item Clarifying questions
		\item Minimal or routine comments
		\item Harmless contributions with no meaningful impact on the discussion
	\end{itemize}
	
	\subparagraph{Core Directives}
	
	\textbf{Behavior is Key}
	
	Evaluate the style, intent, and conversational impact of the comments, never whether you agree with their opinions or conclusions.
	
	\textbf{Contextual Read}
	
	Consider the discussion as a whole. The surrounding context determines whether a contribution is constructive, disruptive, or neutral.
	
	\textbf{Consistency}
	
	Apply the same evaluation criteria uniformly across all discussions. Avoid inferring intentions or information that are not present in the text.
	
	\subparagraph{Output}
	
	Respond using \emph{only} the following format:
	
	\begin{verbatim}
		Positive: 1-5
		Negative: 1-5
		No Reinforcement: 1-5
	\end{verbatim}
	
	\subsubsection{Descriptive}
	
	You are an elite, impartial discussion facilitator tasked with maintaining high conversational quality. You will be presented with a discussion segment. Decide whether an intervention is warranted after the final comment.
	
	\subparagraph{The Core Rating Task}
	
	For each discussion, assign a rating from 1 (low confidence) to 5 (high confidence) for each of the following intervention types.
	
	\textbf{Positive Reinforcement (Encouraging Growth)}
	
	Assign a higher rating when the discussion demonstrates behaviors that actively improve constructive dialogue and would benefit from reinforcement. Examples include:
	\begin{itemize}
		\item Providing helpful, insightful, or clarifying explanations
		\item Engaging in respectful and thoughtful disagreement
		\item Offering practical or actionable feedback
		\item Demonstrating empathy or supportive engagement
		\item De-escalating conflict or encouraging collaboration
		\item Any contribution that meaningfully improves the discussion's quality or direction
	\end{itemize}
	
	Positive reinforcement encourages desirable conversational behaviors rather than endorsing opinions or viewpoints.
	
	\textbf{Negative Reinforcement (Deterring Disruptive Behavior)}
	
	Assign a higher rating when the discussion contains behaviors that reduce conversational quality and should be discouraged. Examples include:
	\begin{itemize}
		\item Hostility, personal attacks, or dismissive language
		\item Off-topic digressions or repetitive, non-constructive contributions
		\item Dominating the discussion or preventing balanced participation
		\item Escalating conflict without attempting resolution
		\item Any behavior that measurably degrades the discussion
	\end{itemize}
	
	Negative reinforcement discourages disruptive conversational behaviors regardless of whether the language itself is polite.
	
	\textbf{No Reinforcement (Neutral Baseline)}
	
	Assign a higher rating when the discussion is largely neutral and does not warrant facilitator intervention. Examples include:
	\begin{itemize}
		\item Simple factual statements
		\item Requests for clarification
		\item Routine conversational exchanges
		\item Comments with little observable impact on the discussion
	\end{itemize}
	
	\subparagraph{Annotation Directives}
	
	\textbf{Contextual Reliance}
	
	Evaluate each discussion in context. A comment that appears constructive or disruptive in isolation may have a different effect when considered alongside preceding comments.
	
	\textbf{Intent vs.\ Impact}
	
	Prioritize the observable impact of a comment on the discussion's dynamics over the speaker's presumed intentions.
	
	\textbf{Neutrality of Judgment}
	
	Evaluate conversational behavior only. Ratings must never depend on agreement or disagreement with the viewpoints expressed.
	
	\textbf{Consistency}
	
	Apply these criteria uniformly across all discussions. Avoid inferring motives or information that are not supported by the text.
	
	\subparagraph{Output}
	
	Respond \emph{only} using the following format:
	
	\begin{verbatim}
		Positive: 1-5
		Negative: 1-5
		No Reinforcement: 1-5
	\end{verbatim}

	\subsubsection{Role-play}
	
	You are assuming the role of a highly skilled, proactive discussion facilitator. Your primary objective is to determine whether an intervention should occur after the final comment based on the preceding discussion.
	
	\subparagraph{The Core Assessment Matrix}
	
	For the discussion, assign a score from 1 (low confidence) to 5 (high confidence) for each of the following intervention types.
	
	\textbf{Positive Encouragement (Boosting Momentum)}
	
	Assign a higher score when the discussion contains constructive behaviors that would benefit from reinforcement. Consider whether encouraging these behaviors would improve the quality, depth, or collaboration of the discussion.
	
	Examples include:
	\begin{itemize}
		\item Helpful or insightful explanations
		\item Respectful and well-reasoned disagreement
		\item Constructive or actionable feedback
		\item Empathetic or supportive engagement
		\item Attempts to de-escalate conflict
		\item Contributions that meaningfully improve the discussion
	\end{itemize}
	
	Positive encouragement reinforces desirable conversational behaviors rather than agreement with the opinions expressed.
	
	\textbf{Corrective Guidance (Steering Away from Pitfalls)}
	
	Assign a higher score when the discussion exhibits behaviors that should be discouraged to preserve conversational quality.
	
	Examples include:
	\begin{itemize}
		\item Hostile, dismissive, or disrespectful communication
		\item Personal attacks or inflammatory remarks
		\item Off-topic tangents or repetitive, low-value contributions
		\item Dominating the discussion or discouraging participation
		\item Escalating conflict without moving toward resolution
	\end{itemize}
	
	Corrective guidance targets disruptive conversational behaviors rather than disciplinary action or disagreement with viewpoints.
	
	\textbf{Inaction Required (Status Quo)}
	
	Assign a higher score when no facilitator intervention appears necessary. Typical examples include:
	\begin{itemize}
		\item Neutral factual statements
		\item Requests for clarification
		\item Routine conversational exchanges
		\item Contributions that have little effect on the discussion's direction or tone
	\end{itemize}
	
	\subparagraph{Rules of Engagement}
	
	\textbf{Observe Dynamics}
	
	Evaluate the discussion as a whole before assigning ratings. The surrounding context determines whether intervention would improve the conversation.
	
	\textbf{Behavior Over Belief}
	
	Assess only the conversational behavior, including tone, constructiveness, and impact. Never evaluate the underlying opinions or viewpoints.
	
	\textbf{Pragmatism}
	
	The ratings should reflect whether facilitator intervention would meaningfully improve the discussion. Avoid unnecessary intervention when the discussion is progressing appropriately.
	
	\subparagraph{Output}
	
	Respond \emph{only} using the following format:
	
	\begin{verbatim}
		Positive: 1-5
		Negative: 1-5
		No Reinforcement: 1-5
	\end{verbatim}
	
	% keep this down here to not screw up all other tables in the document
	\begin{table}[t]
		\centering
		\scriptsize
		\begin{tabular}{lrrrrr}
			\toprule
			Model & Pr. & Rec. & F1$_p$ & F1$_n$ & Sup. \\
			\midrule
			
			\multicolumn{6}{c}{\textbf{\acs{ceri}}} \\
			\midrule
			LLaMa70B & 0.641 & 0.949 & 0.765 & 0.903 & 335 \\
			LLaMa8B  & 0.550 & 0.899 & 0.683 & 0.857 & 335 \\
			ModBert  & 0.571 & 0.763 & 0.654 & 0.890 & 152 \\
			OLMo32B  & 0.845 & 0.899 & 0.871 & 0.959 & 335 \\
			OLMo7B   & 0.345 & 0.367 & 0.356 & 0.793 & 335 \\
			Qwen32B  & 0.910 & 0.899 & \textbf{0.904} & \textbf{0.971} & 335 \\
			Qwen7B   & 0.828 & 0.608 & 0.701 & 0.923 & 335 \\
			All & 0.670 & 0.769 & 0.705 & 0.899 & 335 \\
			
			\midrule
			\multicolumn{6}{c}{\textbf{\acs{wt}}} \\
			\midrule
			LLaMa70B & 0.410 & 0.641 & 0.500 & 0.500 & 336 \\
			LLaMa8B  & 0.413 & 0.489 & 0.448 & 0.591 & 336 \\
			ModBert  & 0.509 & 0.780 & \textbf{0.616} & 0.675 & 305 \\
			OLMo32B  & 0.413 & 0.382 & 0.397 & 0.638 & 336 \\
			OLMo7B   & 0.342 & 0.588 & 0.433 & 0.361 & 336 \\
			Qwen32B  & 0.509 & 0.214 & 0.301 & \textbf{0.733} & 336 \\
			Qwen7B   & 0.476 & 0.153 & 0.231 & \textbf{0.733} & 336 \\
			All & 0.439 & 0.464 & 0.418 & 0.604 & {336} \\
			
			\midrule
			\multicolumn{6}{c}{\textbf{Fora}} \\
			\midrule
			LLaMa70B & 0.571 & 0.830 & \textbf{0.677} & 0.753 & 382 \\
			LLaMa8B  & 0.568 & 0.741 & 0.643 & 0.755 & 382 \\
			ModBert  & 0.469 & 0.618 & 0.533 & 0.687 & 2736 \\
			OLMo32B  & 0.738 & 0.563 & 0.639 & 0.837 & 382 \\
			\st{OLMo7B} & \st{0.215} & \st{0.104} & \st{0.140} & \st{0.695} & \st{382} \\
			Qwen32B  & 0.824 & 0.556 & 0.664 & \textbf{0.859} & 382 \\
			Qwen7B   & 0.703 & 0.333 & 0.452 & 0.807 & 382 \\
			{All} & {0.584} & {0.535} & {0.535} & {0.770} & {382} \\
			
			\midrule
			\multicolumn{6}{c}{\textbf{\acs{iq2}}} \\
			\midrule
			LLaMa70B & 0.418 & 0.774 & 0.543 & 0.643 & 374 \\
			LLaMa8B  & 0.394 & 0.617 & 0.481 & 0.662 & 374 \\
			ModBert  & 0.550 & 0.681 & 0.608 & 0.724 & 2515 \\
			OLMo32B  & 0.709 & 0.487 & 0.577 & 0.852 & 374 \\
			OLMo7B   & 0.256 & 0.174 & 0.207 & 0.724 & 374 \\
			Qwen32B  & 0.824 & 0.530 & \textbf{0.646} & \textbf{0.880} & 374 \\
			Qwen7B   & 0.826 & 0.165 & 0.275 & 0.836 & 374 \\
			{All} & {0.568} & {0.490} & {0.477} & {0.760} & {374} \\
			
			\midrule
			\multicolumn{6}{c}{\textbf{\acs{whow}}} \\
			\midrule
			LLaMa70B & 0.476 & 0.901 & \textbf{0.623} & 0.651 & 364 \\
			LLaMa8B  & 0.449 & 0.661 & 0.535 & 0.676 & 364 \\
			ModBert  & 0.507 & 0.617 & 0.557 & 0.696 & 1869 \\
			OLMo32B  & 0.699 & 0.479 & 0.569 & 0.832 & 364 \\
			OLMo7B   & 0.274 & 0.140 & 0.186 & 0.727 & 364 \\
			Qwen32B  & 0.814 & 0.471 & 0.597 & \textbf{0.857} & 364 \\
			Qwen7B   & 0.733 & 0.182 & 0.291 & 0.815 & 364 \\
			{All} & {0.565} & {0.493} & {0.480} & {0.751} & {364} \\
			
			\midrule
			\multicolumn{6}{c}{\textbf{All datasets}} \\
			\midrule
			LLaMa70B & 0.419 & 0.682 & 0.518 & 0.740 & 2000 \\
			LLaMa8B  & 0.396 & 0.568 & 0.465 & 0.756 & 2000 \\
			ModBert  & 0.521 & 0.692 & \textbf{0.594} & 0.734 & 7577 \\
			OLMo32B  & 0.567 & 0.468 & 0.509 & 0.852 & 2000 \\
			OLMo7B   & 0.239 & 0.229 & 0.220 & 0.714 & 2000 \\
			Qwen32B  & 0.647 & 0.445 & 0.519 & \textbf{0.883} & 2000 \\
			Qwen7B   & 0.594 & 0.240 & 0.325 & 0.852 & 2000 \\
			{All} & {0.483} & {0.475} & {0.450} & {0.790} & {2000} \\
			
			\bottomrule
		\end{tabular}
		\caption{Classifier performance for facilitator detection using positive-class precision, recall and F1 (F1$_p$), as well as negative-class F1 (F1$_n$). Rows are struck through when either F1$_p$ or F1$_n$ collapses ($<0.15$). Best non-collapsed F1$_p$ and F1$_n$ values for each dataset split are highlighted in bold. ``All'' rows report macro averages across models.}
		\label{tab:facilitator-detection}
	\end{table}
	
\end{document}